## Title

- A Critical Edge Number Revealed for Phase Stabilities of Two-Dimensional Ball-Stick Polygons

## Authors

Ruijian Zhu[1,2], Yanting Wang[1,2,*]

## Affiliations

[1] CAS Key Laboratory of Theoretical Physics, Institute of Theoretical Physics, Chinese Academy of Sciences, 55 East Zhongguancun Road, P. O. Box 2735, Beijing 100190, China

[2] School of Physical Sciences, University of Chinese Academy of Sciences, 19A Yuquan Road, Beijing 100049, China

* wangyt@itp.ac.cn

## Abstract

Phase behaviours of two-dimensional (2D) systems constitute a fundamental topic in condensed matter and statistical physics. Although hard polygons and interactive point-like particles are well studied, the phase behaviours of more realistic molecular systems considering intermolecular interaction and molecular shape remain elusive. Here we investigate by molecular dynamics simulation phase stabilities of 2D ball-stick polygons, serving as simplified models for molecular systems. Below the melting temperature $T_m$, we identify a critical edge number $n_c = 4$, at which a distorted square lattice emerges; when $n < n_c$, the triangular system stabilizes at a spin-ice-like glassy state; when $n > n_c$, the polygons stabilize at crystalline states. Moreover, in the crystalline state, $T_m$ is higher for polygons with more edges at higher pressures but exhibits a crossover for hexagon and octagon at low pressures. A theoretical framework taking into account the competition between entropy and enthalpy is proposed to provide a comprehensive understanding of our results, which is anticipated to facilitate the design of 2D materials.

## MAIN TEXT

## Introduction

Two-dimensional (2D) systems are one of the most attractive topics in condensed matter physics because of their fundamental and practical importance. (Quasi-) 2D materials such as graphene *(1, 2)* and Moiré superlattice *(3)* exhibit novel physical properties originated from the massless Dirac cones *(4-7)* and the corresponding topological effects *(8, 9)*, including superconductivity, semi-conductivity, and insulation. The study on confined 2D ice is heuristic for understanding a wide range of natural phenomena *(10)*, and the ice surface has already been used for controlling self-assembly processes *(11)* and promoting chemical reactions *(12)*. To gain insights from numerous structures and physical properties, a central focus is to understand the phase behaviours of 2D molecular systems, including phase stabilities, crystalline morphologies, and associated phase transitions.

The phase behaviours in 2D systems are more complicated than in three-dimensional systems *(13)*. First proposed by Landau and Peierls and then proved by Mermin and Wagner *(14, 15)*, a 2D crystal has a quasi-long-range translational order as well as a long-range bond-orientational order at a finite temperature. Unlike the 3D case, these two orders do not have to vary synchronously in 2D, leading to the possible intermediate phase of hexatic (with a six-fold symmetry) *(16)* or tetratic (with a four-fold symmetry) *(17)*, characterized by a quasi-long-range bond-orientational order along with a short-range translational order *(16)*, which has no analogue in 3D. Correspondingly, there are three different melting scenarios in 2D: (1) the KTHNY theory *(16, 18-20)* when crystal melts into hexatic and then into liquid via two continuous phase transitions; (2) the hard-disk-like behaviour *(21)* when crystal melts into hexatic via a continuous phase transition and then into liquid via a discontinuous phase transition; (3) the regular solid-liquid phase transition *(22)* when crystal melts into liquid directly without the appearance of the hexatic phase. Most recently, the simulation on truncated rhombs suggested the possibility of the fourth melting scenario when crystal melts into hexatic via a discontinuous phase transition and then into liquid continuously *(23)*.

Realistic systems contain complex intermolecular interactions and molecular shapes, which are known to have dramatic influence on the 2D melting scenario, exceeding the capability of analytical calculations *(24)*. Thanks to the fast development of simulation tools and data-analysis methods, complex model systems are possible to be investigated extensively. Roughly speaking, interaction leads to the enthalpy effect while shape is directly related to the entropy effect. Therefore, complex phase behaviours in 2D material systems may be analyzed from the viewpoint of the competition between entropy and enthalpy.

The entropy-limited case was investigated by studying hard polygons with only the volume-repulsive interaction (see e.g., Refs. *17, 23, 25-28*). A comprehensive simulation study by J. A. Anderson et al. *(25)* provides a complete picture of the melting scenarios of hard regular polygons, suggesting that triangle, square, and hexagon follow the KTHNY theory, pentagon follows a simple solid-liquid transition, and polygons with 7 or more edges follow the hard-disk-like behaviour. They have also simulated a space-filling pentagon to clarify that it is the orientational entropy and the confliction of geometrical symmetry between liquid and crystal, rather than the space-filling, that determine the melting scenario. The colloidal experiments on hard polygons have observed more solid structures, e.g., the rhombic crystal for squares *(29)*, the frustrated glassy state for pentagons *(30)*, and the rotator crystal phase for hexagons *(31)*, some of which can be observed in the simulations of corresponding round-corner polygons *(31, 32)*.

The enthalpy-limited case was studied by simulating various interactive point systems *(33-38)* and performing experiments with confined monolayer colloidal spheres *(39)*. It has been shown that the melting scenario can be tuned by varying the interaction parameters. One of the general conclusions is that there can exist different melting scenarios in the same system at different densities, which has been found in different interactive point-like particles and polygons composed of attractive Lennard-Jones (L-J) beads *(36, 38, 40)*.

Only in recent years did researchers pay attention to more realistic interactive polygons considering both interaction and shape, which leads to qualitatively different phase behaviours. The simulations on L-J beads polygons *(40)* (square, pentagon, and hexagon) reveal that a wide solid-liquid coexistence region similar to the L-J point system emerges at low temperatures, while the melting scenario of corresponding hard polygon is followed at high temperatures. Different melting scenarios under various thermal conditions provide a basic understanding on the thermodynamic competition between two factors: Enthalpy plays a major role at low temperatures/pressures while entropy is dominant at high temperatures/pressures. Experiments on millimetre-sized polygons (triangle and square) with isotropic magnetic repulsion *(41)* exhibit crystalline structures incompatible with the symmetry of monomer shape, e.g., the appearance of the triangular lattice structure for the square system, which results from the confliction between the isotropic interaction and the anisotropic shape.

Since these models are not suitable for modelling real molecular systems, the above results are still inadequate for understanding most 2D materials, in which novel phase behaviours may emerge due to the complex competition between entropy and enthalpy. The ball-stick polygons with an L-J ball on each vertex and adjacent balls connected by valence bonds are ideal for extending previous theoretical studies towards the understanding on real 2D materials: On the one hand, the ball-stick models mimic the molecular shape along with the intermolecular interaction; on the other hand, they can be directly compared with the well-studied model systems we described above. Because the L-J balls are on the vertices, although the anisotropies induced by shape and interaction of each molecule belong to the same symmetry group, they spatially conflict with each other, indicating a more complex competition between entropy and enthalpy, which is closer to realistic systems. Since the confliction is weaker for polygons with more edges where both the shape and interaction are more isotropic, the phase behaviours of polygons with fewer edges are expected to deviate more from the corresponding hard polygons. Moreover, no previous studies have compared phase stabilities between different polygons, so some fundamental questions, such as which physical factor is dominant for phase stabilities under different thermal conditions and how it relates to monomer properties, remain unanswered. Here and below, both monomer and molecule refer to one ball-stick polygon.

In this work, we investigate phase stabilities of 2D regular ball-stick polygons (triangle, square, pentagon, hexagon, and octagon) by molecular dynamics (MD) simulation supplemented with some numerical calculation methods. As summarized in Fig. 1, with increasing edge number, we discover a critical edge number $n_c = 4$ for the phase stabilities in this polygonal family below melting points: The triangular system is stabilized in a spin-ice-like glassy state rather than a crystalline structure at any finite temperatures before melting into liquid via a discontinuous phase transition; the square system exhibits a distorted square lattice structure, featured by a self-similar translational symmetry with a high concentration of topological defects; the pentagon, hexagon, and octagon systems are stabilized with the normal crystalline structures analogous to corresponding hard polygons. Moreover, for the crystalline state, $T_\mathrm{m}$ is higher for polygons with more edges at a higher pressure but exhibits a crossover for hexagon and octagon at a lower pressure. Detailed analysis suggests that the perfect match of interaction and shape in the hexagon system induces an ultra-stability at low pressures, revealing the entropy-dominant mechanism at a higher pressure and the enthalpy-dominant mechanism at a lower pressure.

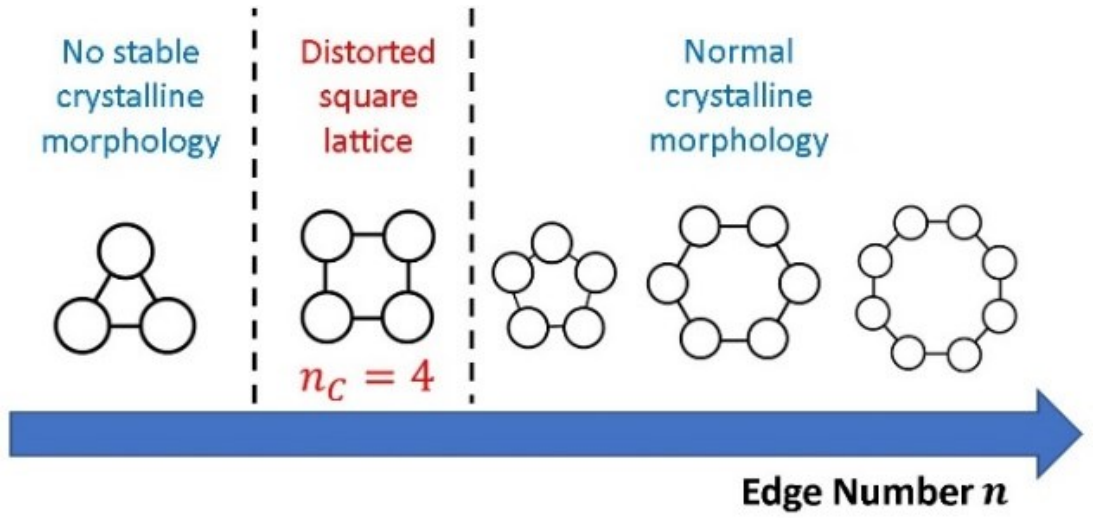

Fig. 1. Illustration of the crystalline phase stabilities of the 2D ball-stick polygons with different edge numbers (*n*). As *n* increases, there exists a critical edge number $n_c = 4$ for the stability of the crystalline morphology.

## Results

### Triangle: Ground-State Degeneracy

#### *Theoretical Determination of Ground States*

The presence of the short-range L-J interaction results in two isotopic unit-cell dimer structures with exactly the same potential energy, splitting from the original one for the dense packing scheme of the hard regular triangle, as illustrated in Fig. 2a. Either type of the isotopes can be spatially duplicated to form the crystalline morphology in Fig. 2b, in which each unit-cell is composed of two neighbouring monomers coloured differently according to their body-orientations, as shown in the upper part of Fig. 2c. The spatial arrangement of unit-cells is shown in the lower part of Fig. 2c, which is a rhombic crystal phase *(29, 32)* ($\alpha_1 \neq \alpha_2$, $\alpha_2$ is smaller than $\pi/2$ but larger than $\pi/3$) different from the crystalline structure of the hard triangular system *(25)* due to the existence of the L-J balls.

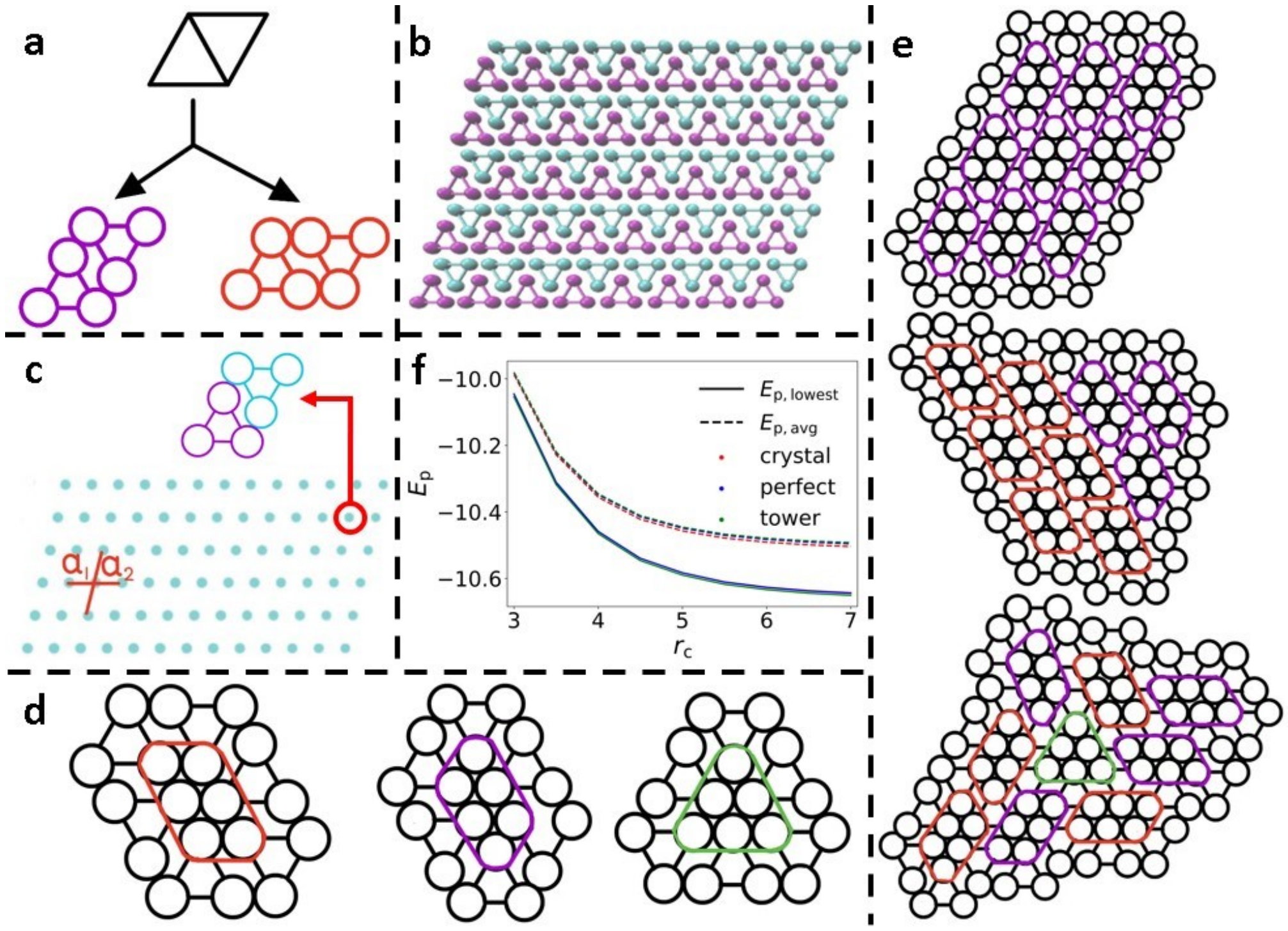

Fig. 2. Ground–state degeneracy of the triangular ball-stick system. **a** Two isotopic structures of the unit-cell, splitting from the original space-filling scheme for the hard triangle. **b** Crystalline morphology. **c** Spatial arrangement of unit-cells. One of $\alpha_1$ and $\alpha_2$, which are the angles formed by two unit-cell primitive vectors, is larger than $\pi/2$, corresponding to a rhombic crystal structure. Each point represents a unit-cell composed of two monomers with opposite body-orientations, as shown at the top of this picture. **d** Three stable local structures composed of the two isotopes with the local clusters in the three configurations all composed of six atoms, which are circled by red, purple, and green frames, respectively. **e** Three typical small hand-made structures constructed based on the three local structures in **d**. The top one is a part of crystal, noted as 'crystal', the middle one is a mixture of the first two local configurations in **d**, noted as 'perfect-mixing', and the bottom one is formed by mixing all the three local configurations in **d**, noted as 'tower-like'. The local clusters are framed with the same colour as in **d**. **f** Potential energies ($E_\mathrm{p}$) of different structures under various cutoff radii ($r_\mathrm{c}$) with the solid lines representing $E_\mathrm{p,lowest}$ and the dashed lines representing $E_\mathrm{p,avg}$. Physical quantities are expressed in L-J units.

These two isotopes can also mix with each other arbitrarily, forming an amorphous state without any geometrical conflicts. In order to have a comprehensive understanding on all the global structures, it is necessary to enumerate the possible local configurations at first. As shown in Fig. 2d, the stacking of the two types of isotopes can form three non-equivalent local clusters. The exhaustive geometrical enumeration (details in SI) suggests that the first two local configurations in Fig. 2d can fill in the whole space by their own to construct a crystalline structure or by mixing with the other one to form an amorphous state, while the third one cannot fill in the whole space by itself and must mix with the first two to avoid the creation of geometrical defects.

If we count in only the L-J interaction from nearest-neighbouring atoms, all the global structures without geometrical conflicts, including both crystalline and amorphous ones, have exactly the same potential energy when the system is so large that the energy cost on the boundary can be ignored, attributed to the fact that the atom clusters in all three of the above stable local structures, which are framed respectively by red, purple, and green in Fig. 2d, have 9 atom pairs. Therefore, a huge number of possible global structures with identical potential energy are all the ground states at $T = 0$, i.e., the triangular system has a ground-state degeneracy, which is very similar to the ground-state degeneracy in the spin-ice system *(42)*, where the oxygen atoms adopt a fixed lattice arrangement while hydrogen atoms can be either ordered or disordered without altering the potential energy.

When the interaction exceeds nearest neighbours, the potential energy is difficult to be analytically calculated, so we instead calculate it numerically. We first construct three typical structures with 30 monomers, as shown in Fig. 2e, where the top one composed of solely the first local structure in Fig. 2d, the middle one containing both the first and the second structures, and the bottom one formed by mixing all the three local structures together. For comparison, we calculate two quantities related to the potential energy $E_{\mathrm{p}}$ for the three structures in Fig. 2e: (1) $E_{\mathrm{p,avg}}$, the potential energy averaged over nine clusters, as circled in each structure; (2) $E_{\mathrm{p,lowest}}$, the lowest potential energy of the clusters. The first one quantifies the global property while the second one provides the information on the local stability. By combining these two quantities, we can have a good estimation on the ground state(s). These two quantities are calculated with the cutoff radius ($r_{\mathrm{c}}$) ranging from $3\sigma$ which is approximately the size of the atom cluster, to $7\sigma$ at which the L-J interaction has a very small value about $10^{-5}$, where $\sigma$ is the van der Waals (vdW) diameter, and the results are shown in Fig. 2f. These three structures have nearly the same potential energy, as the difference of the potential energy varies with $r_{\mathrm{c}}$ but remains always on the order of $10^{-4}$. Therefore, we affirm that the ground state in the triangular system is degenerate, with two crystalline morphologies purely formed by one of the isotopes and a huge amount of amorphous configurations formed by mixing different local structures in Fig. 2d. At any finite temperatures, the vast amount of possible ways to mix the isotopes results in a very large configurational entropy, which allows the amorphous state to have a lower free energy than the crystal state.

***Confirmation of the Ground-State Degeneracy by Simulation***

To verify the above structural origin of the crystalline instability, we performed a replica exchange MD (REMD) simulation on 625 ball-stick triangles with 26 temperature replicas ranging from 2.55 to 6.1. An energy minimization on the final structure of the replica at the lowest simulated temperature has been performed after the REMD simulation, which is theoretically the most stable configuration at low temperatures *(43)*. All the three local structures in Fig. 2d appear in the optimized structure, whose representative snapshot is shown in Fig. 3a. Moreover, benefiting from the fact that the bond length ($1.5\sigma$) is just a little larger than the equilibrium distance of $\sqrt[6]{2}\sigma$, at finite temperatures, there exist some 'meta-space-filling' structures (coloured in blue), where at least two atoms in the same cluster are connected by bond directly, leading to small geometrical defects and a little higher potential energy than the space-filling ones at zero temperature. As a result, they should vanish at $T=0$ but could appear at a slightly higher temperature for they further enlarge the configurational entropy.

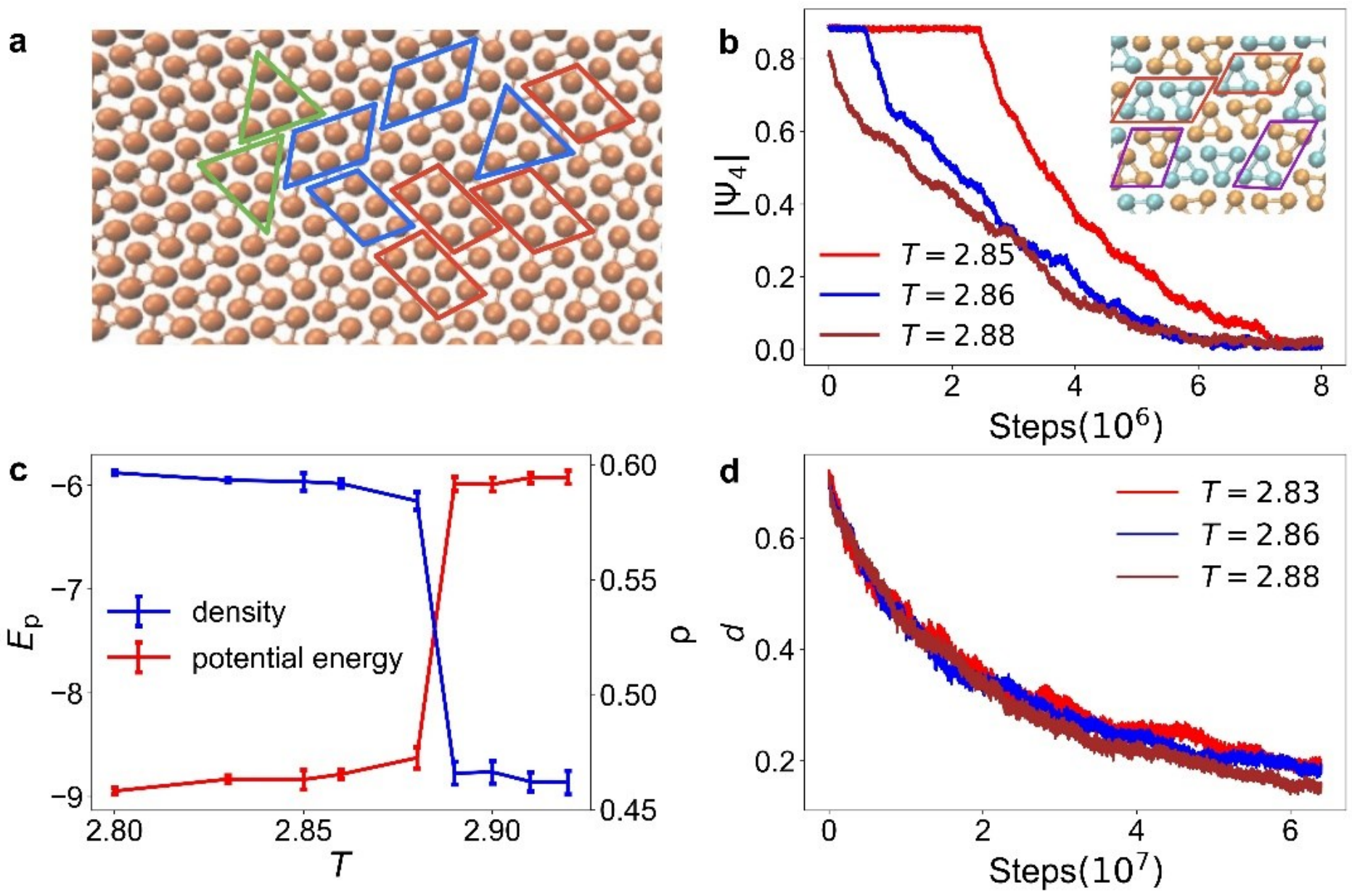

**Fig. 3 Simulation evidence for the ground-state degeneracy. a** A small part of the representative snapshot for the most stable structure determined by the replica exchange molecular dynamics simulation. The green and red frames mark different local configurations in Fig. 2d, and the blue frames illustrate the existence of meta-space-filling structures at finite temperatures. Red and green have the same meaning as in Fig. 2d, while we do not distinguish red and purple here. **b** Amorphization process characterized by the decay of bond-orientational order parameter $|\Psi_4|$ at three temperatures below the melting point. The inset is a small part of the snapshot at *T* = 2.86 after amorphization. **c** Caloric curve (potential energy $E_p$ versus temperature *T*) and density curve (density $\rho$ versus *T*), both with the discontinuous changes located at *T* = 2.88–2.89. The error-bars represent standard deviations, resulting from the thermal fluctuations in one simulation trajectory. **d** Time evolution of the unit-cell order parameter *d* in the simulation starting from a locally disordered crystalline structure at *T* = 2.83, 2.86, and 2.88. Each curve plotted in **b** and **d** is based on the data from a single simulation. Due to the randomness of the dynamical process, the starting point and the rate of amorphization may take different values in different trajectories. Physical quantities are expressed in L-J units.

As the amorphous state is more stable than crystal at finite temperatures, we expect that, driven by entropy, the crystal state will spontaneously destabilize into the amorphous state. To check this, MD simulations starting from crystalline morphology are performed at pressure $P=0$ and various temperatures. As quantified by the bond-orientational order parameter $\Psi_4$ (see Methods for the definition) plotted in Fig. 3b, which initially keeps a high value (larger than 0.8) corresponding to the rhombic crystal state *(29, 32)* and then decays after certain simulation steps until a very low value due to its amorphous nature, the amorphization takes place at $T \geq 2.85$, lower than the melting point $T = 2.88 \sim 2.89$ manifested by the sharp changes in the caloric and density curves plotted in Fig. 3c. The non-equilibrium amorphous state whose potential energy and density are close to those of crystal is featured by the mixing of two isotopes, as shown in the inset of Fig. 3b, in agreement with our theoretical prediction described above. At lower temperatures, the fact that amorphization does not occur within the finite simulation time should be attributed to the existence of relatively high energy barrier(s) between the crystalline and amorphous states. To verify this, we manually introduce a very small disordered region (about 30 monomers) into the crystalline configuration (see Methods for technical details) and run *NPT* simulations at $T = 2.83$ , 2.86, and 2.88, respectively. The evolution of the two monomers in each unit-cell is traced by the unit-cell order parameter defined as $d(t) = \frac{1}{n} < \sum_{\text{unit-cell}} \Theta(l_c - l(t)) >$, where the number of unit-cells *n* equals to half of the number of monomers, *l*(*t*) is the distance between the centre-of-masses (COMs) of the two monomers originally in the same unit-cell at time *t*, $l_c$ =5.77 is the cut-off value of *l*, and $\Theta$ is the Heaviside step function. This order parameter quantifies the fraction of remaining unit-cells at time *t*, taking a value close to 1 in crystal and 0 in liquid. The time evolution of $d$ in $6.4\times10^7$ steps after a relaxation of $4\times10^6$ steps is shown in Fig. 3d, demonstrating that, once the energy barrier is overcome, the system will rearrange into the amorphous state continuously. We expect that *d* would finally drop to approximately zero after sufficiently long time, as a result of reorganization of the whole system. Although regular MD simulations at much lower temperatures would suffer from very slow dynamics, we have managed to verify that the amorphous state is still more stable than the crystalline state even at a temperature as low as *T* = 1.5 by running an additional REMD simulation focusing on the low-temperature region (see Fig. S4 in SI). Moreover, we never observe an inverse pathway from the amorphous state back to the crystal, as appeared in the simulation for the hard-triangle system *(25)*, in agreement with the fact that the amorphous state is more stable than the crystal benefiting from the large configurational entropy.

### *Glassy Nature of the Spin-Ice-Like Amorphous State*

Detailed analysis on dynamics provides a more comprehensive understanding on this amorphous state. As shown in Fig. 4a, in the amorphous state at $T = 2.83$, 2.86, and 2.88, the calculated mean-squared displacements (MSDs) $\Delta^2$ of monomers grow in a strongly frustrated way, clearly distinguished from the one for crystal or liquid shown in Fig. S5. We further plot the non-Gaussian parameter for MSD in Fig. 4b, defined as $\alpha = \frac{3 < (r(t) - r(0))^4 >}{5 < (r(t) - r(0))^2 >^2} - 1$, which is expected to be 0 for the Brownian motion and larger than 0 when some monomers move faster than others. The calculated $\alpha$ for this amorphous state is higher than in the crystal or liquid state, indicating a strong dynamical heterogeneity and the fact that the amorphous state is a glassy state from the dynamical perspective *(44)*.

Overall, considering both the structural and dynamical characteristics, the 2D ball-stick triangular system is in a spin-ice-like glassy state at any finite temperatures below melting point *(45)*, as a result of the strong confliction between interaction and shape. As shown in Fig. 4c, each stable local atom cluster has six neighbouring clusters and contains six atoms with four close to each other while the other two away from them, so our system shares some common features with a spin-ice system defined on a triangular lattice with the '4-in-2-out' ice rule *(46)*. However, the ball-stick triangular system is more complicated than the spin-ice system due to the following two features: (1) Not only the pair interaction from nearest neighbours is considered; (2) The triangular lattice is defined topologically without well-defined lattice structure (no on-site oxygen atoms), so the spatial structure is totally disordered.

The detailed examination on Fig. 3a and Fig. 3b suggests that no matter the isotopes mix or not, each atom has six atomic nearest neighbours arranging in a nearly dense-packing way. This local structure is shown clearer in the inset of Fig. 4d, where the atom shadowed in purple has six neighbours circled by purple frames. Inspired by the spin-glass system *(47, 48)* whose magnetization order parameter characterizes the order-disorder transition while the Edwards-Anderson order parameter measures the glass-liquid transition, here it is also possible to find an order parameter other than $\Psi_4$ to distinguish the glassy state from the liquid one. We therefore employ the bond-orientational order parameter $\Psi_{6,\mathrm{atom}}$ associated with all atoms indistinguishably of the presence of bonds instead of molecular COMs. The discontinuous drop of $\Psi_{6,\mathrm{atom}}$ at $T = 2.88\sim2.89$ in Fig. 4d demonstrates that the spin-ice-like state melts into liquid via a first-order phase transition, in agreement with the caloric and density curves in Fig. 3c.

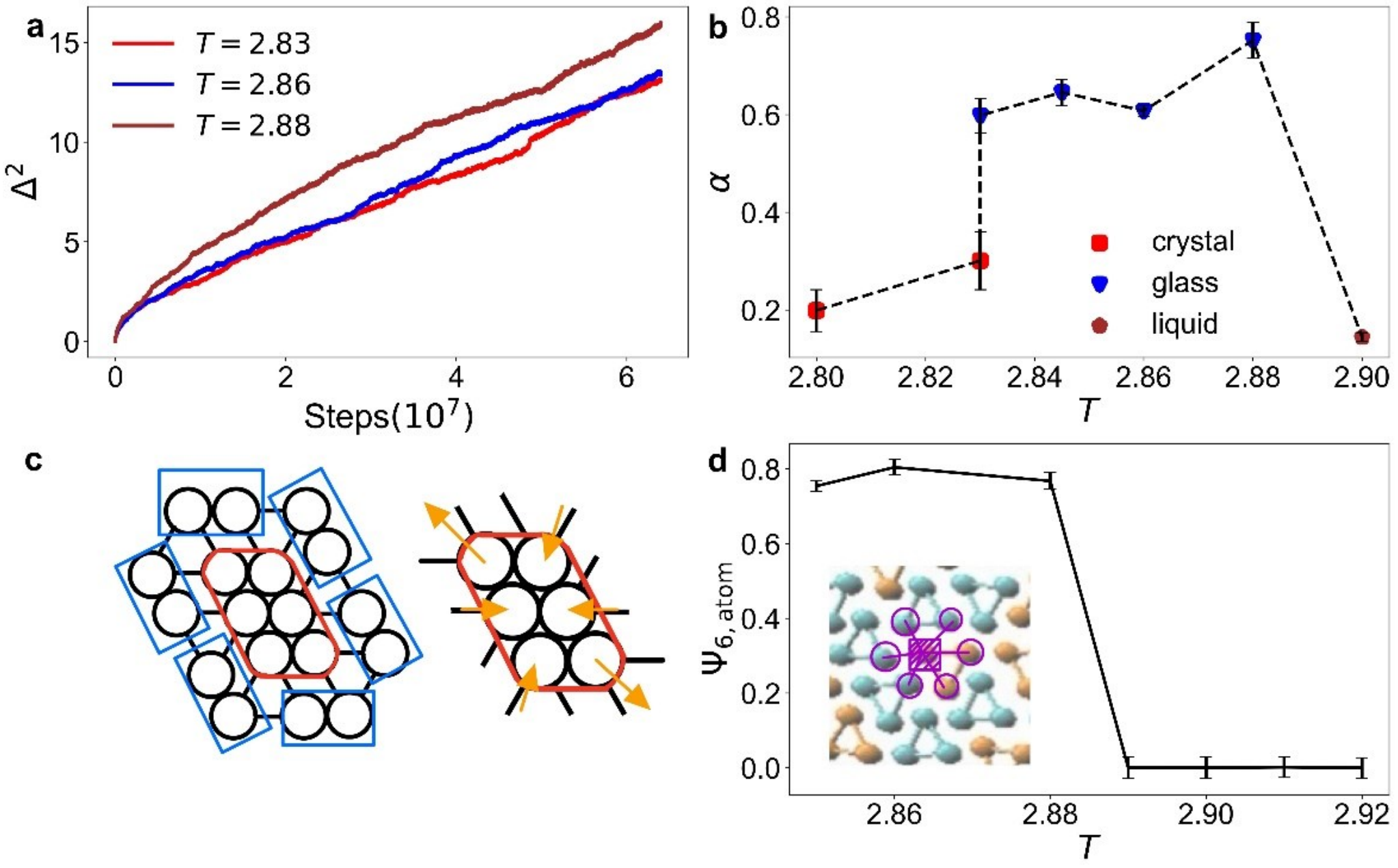

**Fig. 4. Physical nature of the non-equilibrium amorphous state and the phase transition.** **a** Mean-squared displacements (MSDs) ($\Delta^2$) in the glassy state exhibiting frustrated diffusive behaviour. **b** Non-Gaussian parameters for MSD $\alpha$ vs. temperature *T*. The two values at *T* = 2.83 correspond to the crystalline state (smaller one) and the glassy state (larger one). **c** Schematic illustration of the spin-ice-like state defined on a topologically triangular lattice with the '4-in-2-out' ice-rule. The left panel shows the six neighbouring atom-clusters (circled in blue) of the central cluster (circled in red) while the right one illustrates that the six atoms in the cluster exhibit the '4-in-2-out' feature along the direction indicated by the orange arrows. **d** Bond-orientational order parameter $\Psi_{6,\mathrm{atom}}$ vs. *T*, whose discontinuous change suggests a first-order nature of the melting of the spin-ice-like state. The inset is a small portion of the snapshot exhibiting the local packing scheme, where the purple lines and circles highlight the six neighbouring atoms of the central atom shadowed in purple arranged in a nearly dense-packing way. The error-bars in this figure represent standard deviations, resulting from the thermal fluctuations in one simulation trajectory. Physical quantities are expressed in L-J units.

Furthermore, the MD simulation at *P* = 10 (details in SI) manifests that the crystalline morphology rearranges into a spin-ice-like state at finite temperatures and the melting transition from spin-ice-like to liquid is still a first-order phase transition, indicating that the revealed phase behaviour of the 2D ball-stick triangular system is robust with respect to different thermal conditions.

### Square: Distorted Square Lattice

Based on the result from REMD simulation and the consideration on symmetry, the initial state was established as a square lattice shown in Fig. 5a, where the body-orientation of each monomer is not perpendicular to the primary vector of lattice to ensure a higher packing fraction. MD simulations at various temperatures were performed at $P = 0$. The apparent discontinuous changes in the caloric curve, density curve, and $\Psi_4 - T$ plotted in Fig. 5b manifest a first-order phase transition from solid to liquid with the melting point located between 2.92 and 2.93.

The visual examination of Fig. 5c tells us that, in the solid state, although each monomer has four neighbours aligned nearly orthogonally, also reflected by the diffraction pattern shown in Fig. S7a, it apparently distorts from the normal square lattice. This is evidenced by the fact that the Voronoi cells are close to squares but mostly have 6 edges. Moreover, the concentration of topological defects, defined as the fraction of Voronoi cells with more or less than 6 edges, is as high as about 0.42 (data in Fig. S8b), significantly larger than the typical value of crystal (about 0.05) and even hexatic (about 0.123) *(25, 49, 50)*. The quantitative characterization of this state exhibits two seemingly contradict facts: (1) The long-range bond-orientational correlation can be retained until melting, indicated by the correlation function in Fig. 5d, as well as the heatmap in Fig. S7c coloured according to the argument of $\psi_{4,i}$, where the whole picture holds almost the same colour, in agreement with its very narrow distribution; (2)The translational correlation function exhibits a decaying behaviour slightly faster than algebraical, as shown in Fig. S8e.

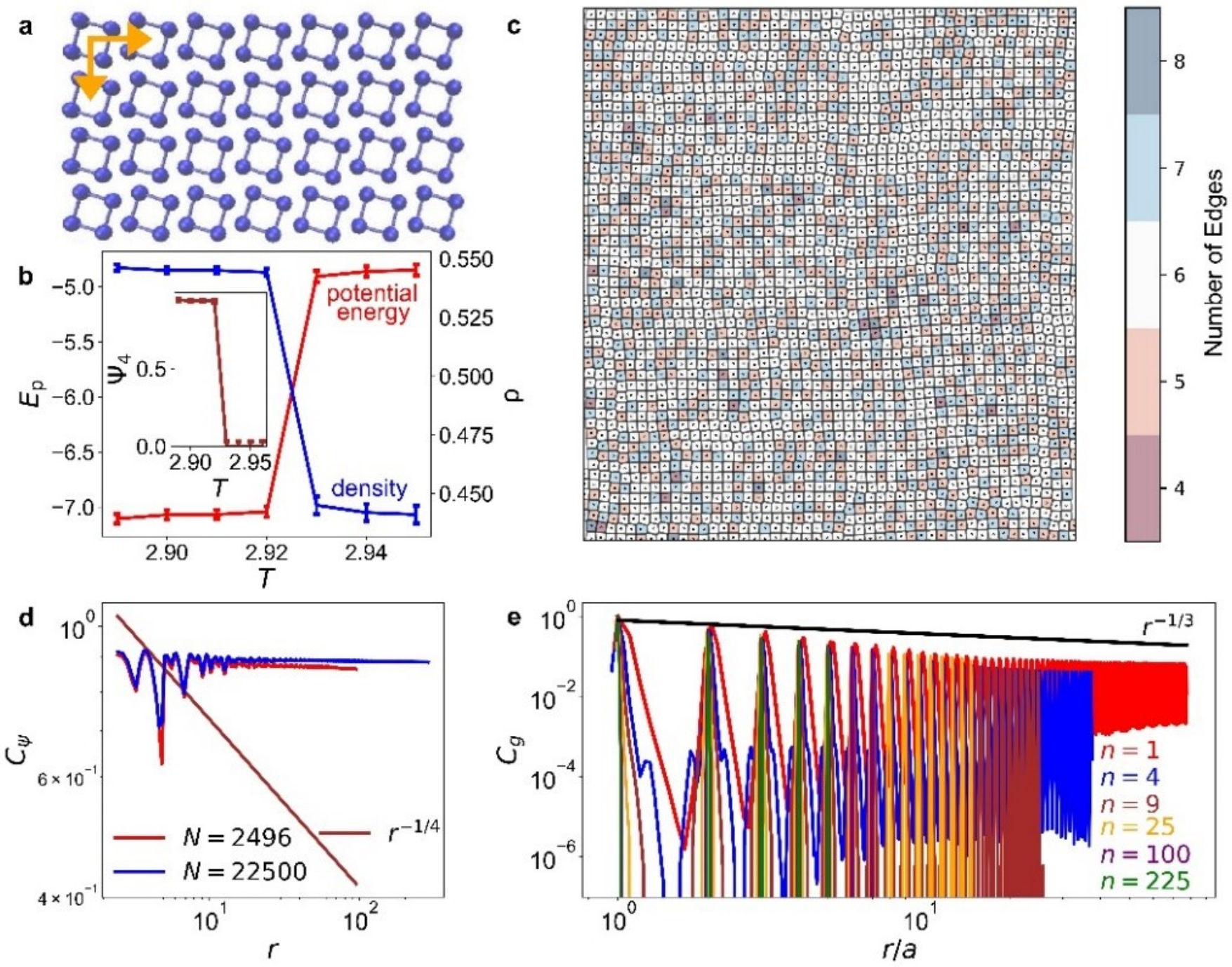

**Fig. 5. Simulation results of the ball-stick square system. a** Initial crystalline morphology. The two orange arrows are the primitive lattice vectors. **b** Caloric curve (potential energy $E_{\mathrm{p}}$ versus temperature *T*) and density curve (density $\rho$

versus *T*). The inset is the bond-orientational order parameter $\Psi_4$ at different temperatures. All the discontinuous changes of these three curves locate at *T* = 2.92~2.93. The error-bars represent standard deviations, resulting from the thermal fluctuations in one simulation trajectory. **c** The spatial arrangement of centre-of-masses (the black dots) at *T* = 2.92. The polygon corresponding to each point is decided by the Delauney tessellation considering the periodic boundary condition and is coloured according to its number of edges. **d** Bond-orientational correlation function (correlation $C_\psi$ as a function of distance *r*) at the melting point for *N* = 2496 monomers and *N* = 22500 monomers, indicating no decay compared to the brown line corresponding to an algebraical decaying behaviour. **e** Translational correlation functions (correlation $C_g$ as a function of *r*) at melting points for *N* = 22500 monomers with different block sizes *n* (number of monomers in each block), decaying faster than algebraically (the black line) locally but slower on the long range. The distance is rescaled by the characteristic length of the coarse-graining *a*. Physical quantities are expressed in L-J units.

To clarify this paradox, we perform simulations on a larger system with *N* = 22500 monomers, whose results are consistent with the above (details in SI). With the larger simulation scale, we identify that, the translational correlation decays relatively fast locally but exhibits a typical algebraical decaying behaviour in the long range, as illustrated by the red line in Fig. 5e, indicating that this solid state is still a crystal. Combined with the above features, we regard it as a distorted square lattice. The translational symmetry is further investigated by a 'renormalization' process, in which we divide the monomers into blocks, with *n* (=1, 4, 9, 25, 100, 225) monomers in each block, and then calculate the translational correlation of the block COMs. After rescaling the distance by the characteristic length of the block, which is just the distance of the first peak in the correlation function, different curves collapse together, as shown in Fig. 5e, exhibiting the so-called self-similarity, which guarantees the existence of the translational symmetry. Furthermore, considering the locally disordered nature, this self-similarity strongly implies that it is a critical point for the ordered and disordered solid states *(51, 52)*, neither totally disordered as the triangular system nor perfectly ordered as a standard crystalline structure for pentagon, hexagon, or octagon. Since this holds at different temperatures (e.g., also *T* = 2.8 in Fig. S8f), the critical point is defined in the shape space in terms of a critical edge number, rather than in the thermal-parameter space.

When cooling down the distorted square lattice, the Voronoi cells have less dispersive areas but surprisingly higher concentrations of topological defects (Fig. S8b and c), even for the case as low as *T* = 0.01. This supports the mechanism that the local distortions are originated from the intrinsic properties of the system, in terms of interaction and shape rather than thermal fluctuations, emphasizing its critical nature in the shape space.

We also perform simulations at *P* = 5 and *P* = 10. Under different pressures, the square system still follows a solid-liquid transition without the appearance of the tetratic phase and the solid structures also share similar features (see SI).

## Pentagon, Hexagon, and Octagon: Entropy-dominant versus Enthalpy-dominant

### *Phase Diagrams*

For these three polygons, the ground-state crystalline structures: striped phase for pentagon, triangular lattice for both hexagon and octagon, can be stabilized until their melting temperatures, attributed to the even weaker confliction between interaction and shape. The snapshots of these crystalline configurations are shown in Fig. 6a, b, and g, respectively. Below we describe the melting scenarios of each polygon under various pressures, and the readers are referred to the Methods section and SI for more details on the calculation procedure and results.

Our MD simulations indicate that the striped phase is the most stable phase at low temperatures for the ball-stick pentagon system, which is also theoretically the closest packing mode for hard pentagon *(53)*. With increasing temperature, the striped phase first experiences a first-order solid-solid phase transition into the rotator crystal phase and then melts into the liquid phase via another first-order phase transition, the same as the melting scenario of hard pentagon *(28)*. In the striped phase, monomers are aligned with two body-orientations to form alternative lines, as shown in Fig. 6a, rendered as two peaks in the body-orientation distribution (the blue line in Fig. 6c). In the rotator crystal phase shown in Fig. 6b, monomers have a totally random body-orientation, demonstrated by the flat distribution of body-orientation (the orange line in Fig. 6c), indicating that this rotator crystal phase is a continuous one *(54)*. The phase behaviour of the pentagon system is summarized in the phase diagram in Fig. 7a, in which the striped phase, rotator crystal phase, and liquid phase appear in sequence as temperature increases at any pressures.

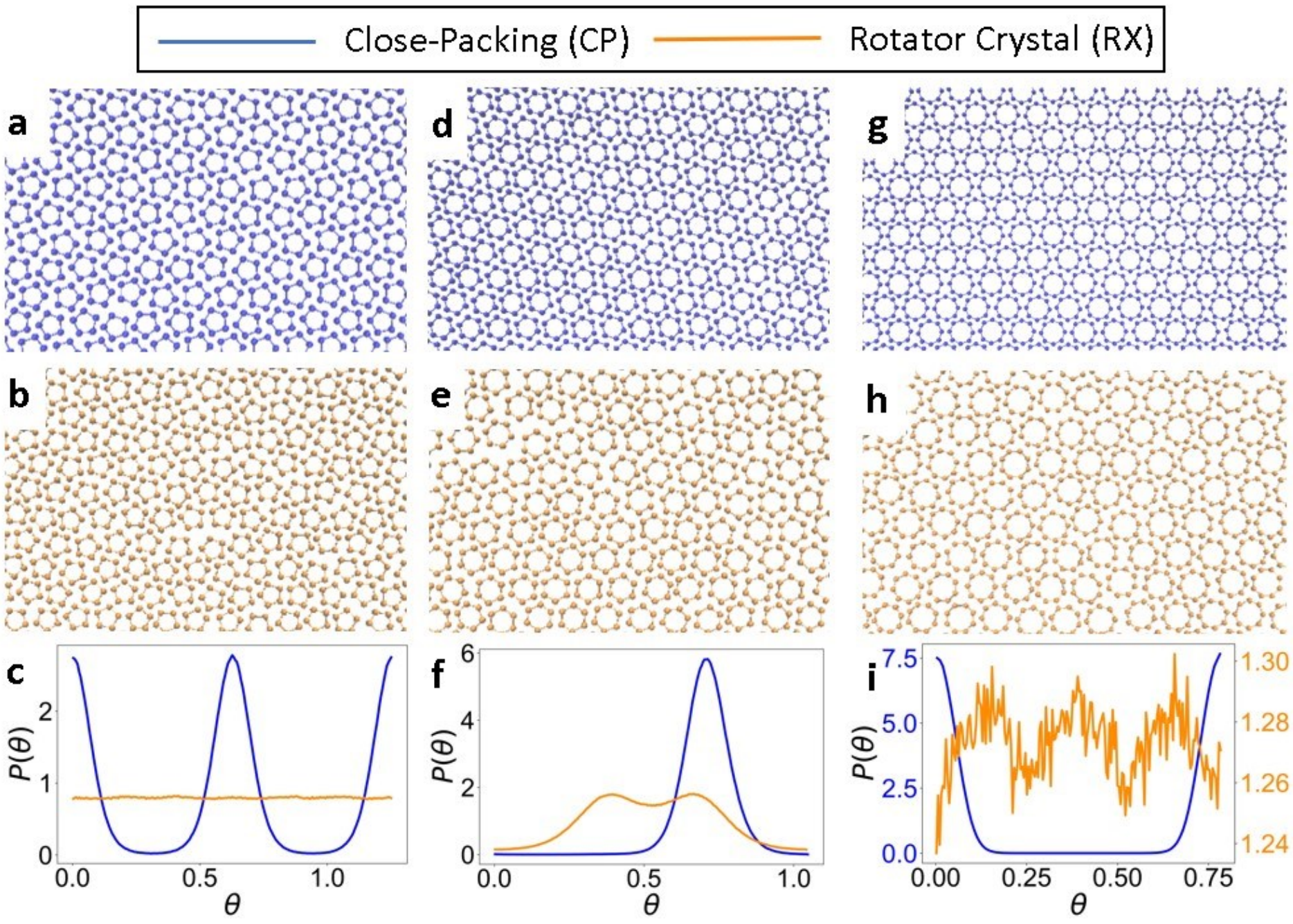

**Fig. 6. Crystalline morphologies of pentagon, hexagon, and octagon systems. a** Striped phase of the pentagon system, in which monomers have two body-orientations comprising alternating lines. **b** Rotator crystal phase of the pentagon system, in which monomers have random body-orientations. **c** Body-orientation distributions for the two crystal phases of the pentagon system, where $\theta$ represents the orientation of monomer with respect to the *x*-axis of the simulation box, ranging from 0 to $2\pi / n$ for *n*-gons, and $P(\theta)$ is the corresponding probability density. **d** Triangular lattice crystal phase of the hexagon system. **e** Rotator crystal phase of the hexagon system, in which monomers have two preferred body-orientations. **f** Body-orientation distributions for the two crystal phases of the hexagon system. **g** Triangular lattice crystal phase of the octagon system. **h** Rotator crystal phase of the octagon system. **i** Body-orientation distribution for the two crystal phases of the octagon system, where the *y* axes are scaled differently to exhibit the trimodal distribution in the rotator crystal phase.

The hexagon system exhibits a rich phase behaviour with increasing pressure. At $P = 0$, the crystalline structure sublimates without the appearance of the liquid state, evidenced by a nearly-zero potential energy and an RDF curve with only one peak shown in Fig. S11 in SI. At $P = 0.5$ and 1.0, the crystalline structure melts into the liquid state via a first-order phase transition. At $P = 1.5$, 2.5, and 5, a rotator crystal phase (Fig. 6e) appears between the crystal state and the liquid state, and both of the two phase transitions are discontinuous. The monomer COMs in the rotator crystal phase form a triangular lattice, but monomers have two body-orientations identified by the two broad peaks in Fig. 6f, so this rotator crystal phase is regarded as a discontinuous one *(54)*. When the pressure increases to $P = 7.5$, the hexatic phase appears between the rotator crystal phase and the liquid phase. In this case, both transitions from the triangular lattice crystal to the rotator crystal and from the hexatic phase to the liquid phase are discontinuous, while the one from the rotator crystal phase to the hexatic phase is continuous, basically following the hard-disk-like behaviour *(21)*. Even at $P = 10$, the melting scenario still follows a hard-disk-like behaviour, instead of the KTHNY theory followed by the hard hexagon or the L-J beads hexagon *(25, 40)*, attributed to the appearance of the rotator crystal phase. The phase diagram is shown in Fig. 7b, in which the rotator crystal phase only appears at relatively high pressures and the hexatic phase emerges at even higher pressures.

The octagon system stabilizes at the triangular lattice up to a finite temperature when it transforms into a rotator crystal phase discontinuously. The rotator crystal of octagon has roughly three preferred body-orientations of monomers, as shown in Fig. 6g and h, similar to the case for hard octagon *(54)*. However, the body-orientation distribution is very rough and the ranges between peaks and valleys are much smaller than that of the discrete rotator crystal phase for hexagon. The melting scenario follows a regular solid-liquid one at relatively low pressures ($P = 0$—5). The hexatic phase appears at higher pressures of $P = 5$—8 and leads to a hard-disk-like behaviour, qualitatively the same as the L-J disk *(40)*. The phase diagram of the octagon system is drawn in Fig. 7c, showing that there are always two solid states, and the hexatic region is very narrow at high pressures and disappears at low pressures.

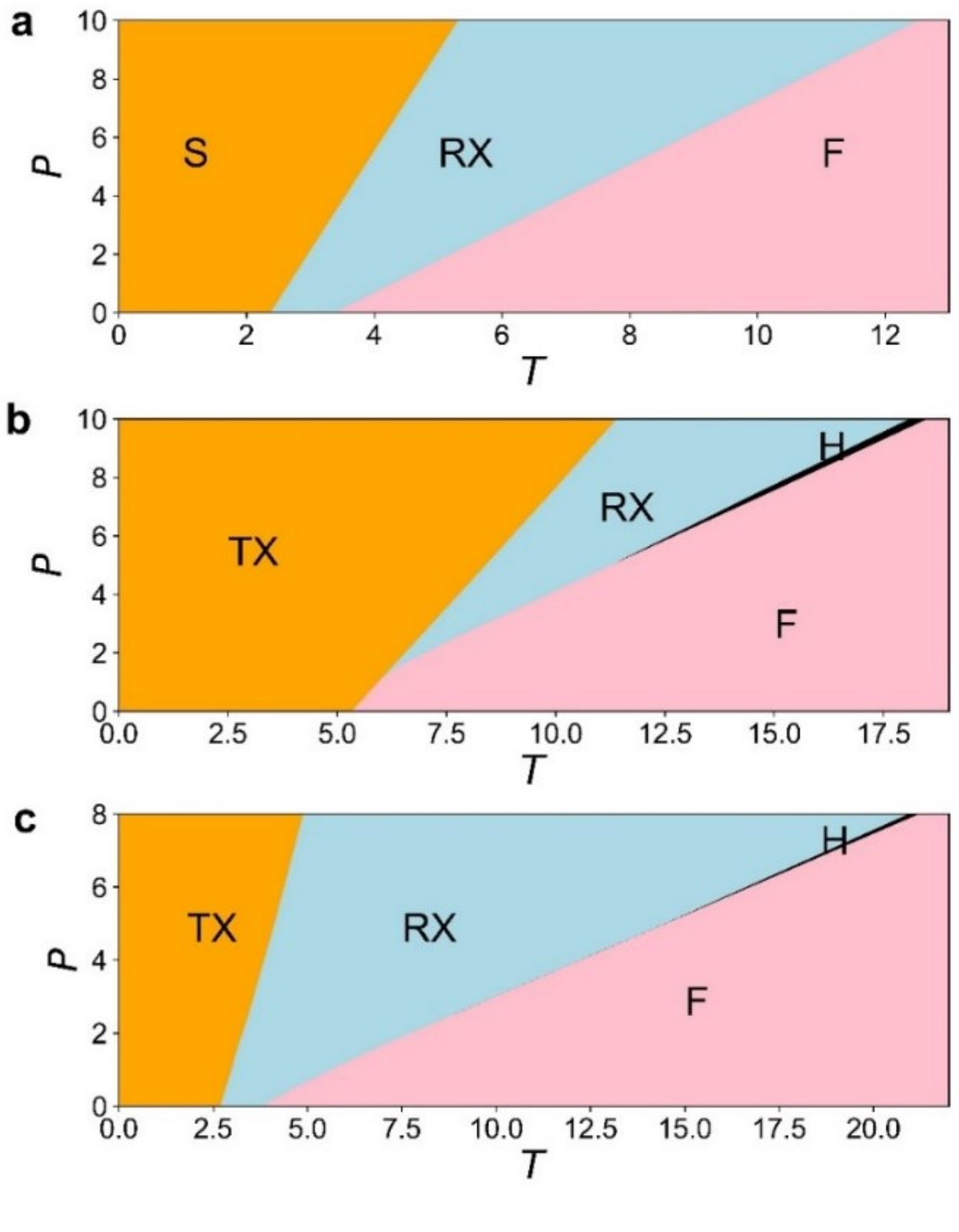

**Fig. 7. Pressure (*P*)-Temperature (*T*) phase diagrams of pentagon (a), hexagon (b), and octagon (c).** The striped phase (S) and the triangular lattice crystal phase (TX) are coloured in orange, the rotator crystal phase (RX) is coloured in light

blue, the hexatic phase (H) is coloured in black (very narrow for hexagon and octagon), and the fluid phase (F) (liquid or gas) is coloured in pink. Physical quantities are expressed in L-J units.

### *Phase Stabilities*

To gain a deeper understanding on the mechanism of phase stabilities, we plot the melting points of different crystalline structures under different pressures in Fig. 8. The melting point increases with pressure and is higher for polygons with more edges at relatively high pressures, which is regarded as the entropy-dominant regime. Meanwhile, the melting-temperature curves for hexagon and octagon have a crossover point located between $P = 1$ and $P = 1.5$. At relatively low pressures ( $P \leq 1$ ), hexagon stays in the triangular-lattice crystal phase before melting into liquid, and has a higher melting point than octagon. By contrast, at relatively high pressures ( $P \geq 1.5$ ), hexagon is in the rotator crystal phase before melting into liquid, and has a melting point lower than octagon. Therefore, the crossover of the melting point locates in the pressure interval where the rotator crystal phase for hexagon appears. Detailed examination on the local structure of triangular-lattice crystal morphology suggests that, as shown in Fig. 6d and much clearer in Fig. 9d, each atom has two neighbouring atoms belonging to two different molecules, and these three atoms form a regular triangular local structure, which is the most stable structure of an L-J cluster composed of three atoms. In other words, the crystal phase for hexagon has the intermolecular interaction perfectly matches the molecular shape, leading to an ultra-stability at low pressures when the phase stability is dominated by enthalpy.

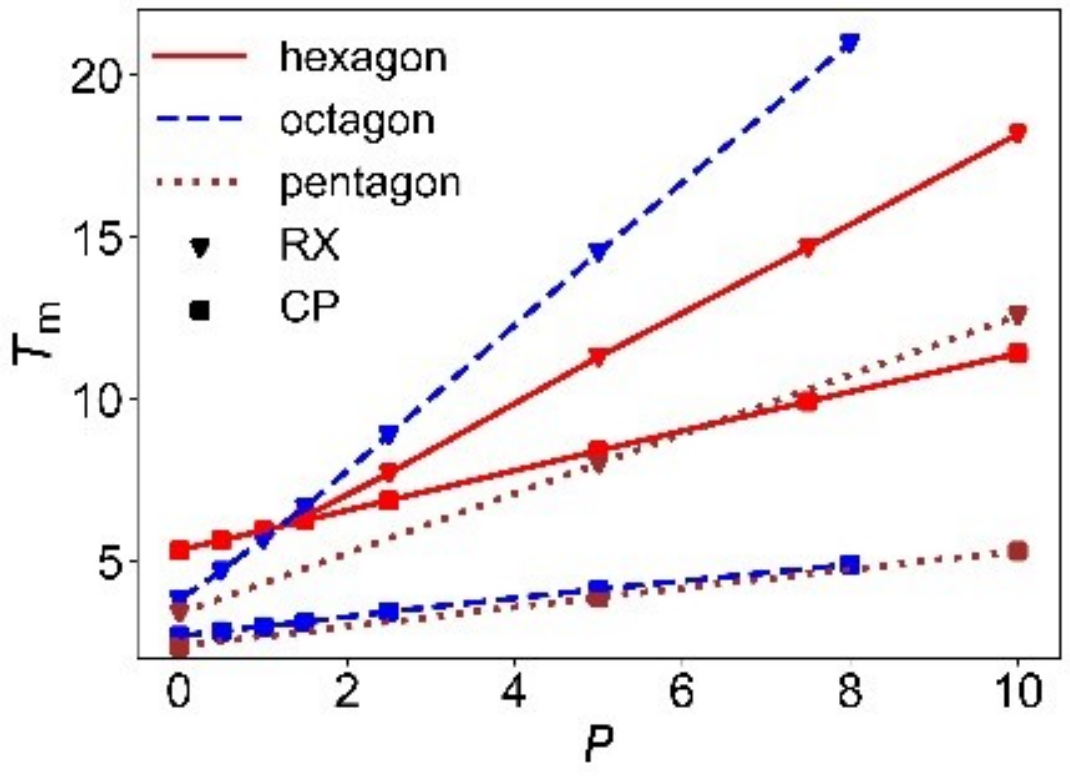

**Fig. 8. Melting point ( $T_m$ ) vs. pressure (*P*).** Lines in different colours represent different polygons, and different symbols represent different solid states: square for close packing (CP) and triangle for rotator crystal (RX). Physical quantities are expressed in L-J units.

The above phenomenon helps us to deepen our understanding on the competition between the entropy and enthalpy. At relatively low pressures, the interaction plays a leading role in phase stability, i.e., whether the local structure is interaction-favoured determines the phase stability, which is regarded as ‘enthalpy-dominant’. Conversely, at relatively high pressures, the molecular shape of a polygon determines its phase stability, corresponding to an ‘entropy-dominant’ mechanism. The striped phase for pentagon and the triangular lattice crystal for octagon do not show the same ultra-stability as hexagon because the interaction and shape there do not perfectly match each other.

## Discussion

In this work, we have investigated phase stabilities of the 2D ball-stick polygonal family (triangle, square, pentagon, hexagon, and octagon) by MD simulation and numerical calculation. We have found a critical edge number of $n_c = 4$ for the crystalline stability of 2D ball-stick polygons. For $n < 4$, the ball-stick triangular system has no stable crystalline structure at any finite temperatures, instead it forms a spin-ice-like state characterized by amorphous structures and glassy-like dynamics due to the ground-state degeneracy originated from the mixing of two types of isotopic dimer structures. For $n = 4$, the ball-stick square system forms a distorted square lattice with a high concentration of topological defects but self-similar translational symmetry below melting. For $n > 4$, the ground-state crystal structures: striped phase for pentagon, triangular lattice for hexagon and octagon, can be stabilized up to the melting temperature. We have also studied the phase stabilities of the crystal states of different polygons by investigating the $P$–$T_m$ curves, and found a melting-point crossover for hexagon and octagon.

All the results in this work may be understood under the theoretical framework based on the competition between entropy and enthalpy. The 2D ball-stick polygons in this study exhibit spatially conflicted anisotropies induced by the molecular shape (entropy) and intermolecular interaction (enthalpy), resulting in a deviation of the stable structures from the space-filling pattern of the corresponding hard polygons, as shown in Fig. 9, in which the dark purple lines represent the connection of COMs of neighbouring polygons determined solely by geometrical consideration, while orange lines represent the case in the ball-stick polygon systems. The critical edge number for the anisotropic confliction is 4: When $n < 4$, the confliction is strong enough to break the crystalline morphology and leads to a spin-ice-like glassy state; when $n > 4$, the confliction is weakened as a perturbation because the shape and interaction of monomers become more isotropic, resulting in phase behaviours close to the corresponding hard polygons; when $n = 4$, the effects of interaction and shape are subtly balanced with each other, served as a critical point in the shape space. For triangle, due to the vanish of ordered structures, no hexatic phase appeared in hard triangular system is observed *(25)*. For square, the distorted lattice directly melts into liquid without going through the tetratic phase appeared in the hard square *(17, 25)* or L-J beads square *(40)* system, attributed to the sufficiently large concentration of topological defects in the solid state. For polygons with more than 4 edges, the appearance (or disappearance) of a hexatic phase is basically the same as corresponding hard polygons *(25)*. Moreover, under various thermal conditions, two different mechanisms for phase stabilities compete with each other: The entropy-dominant mechanism at relatively high pressures results in a higher melting point for polygons with more edges due to the smaller orientational entropy, while the enthalpy-dominant mechanism at relatively low pressures takes interaction details into account, inducing an ultra-stability for hexagon whose crystalline structure has an interaction-favoured local structure.

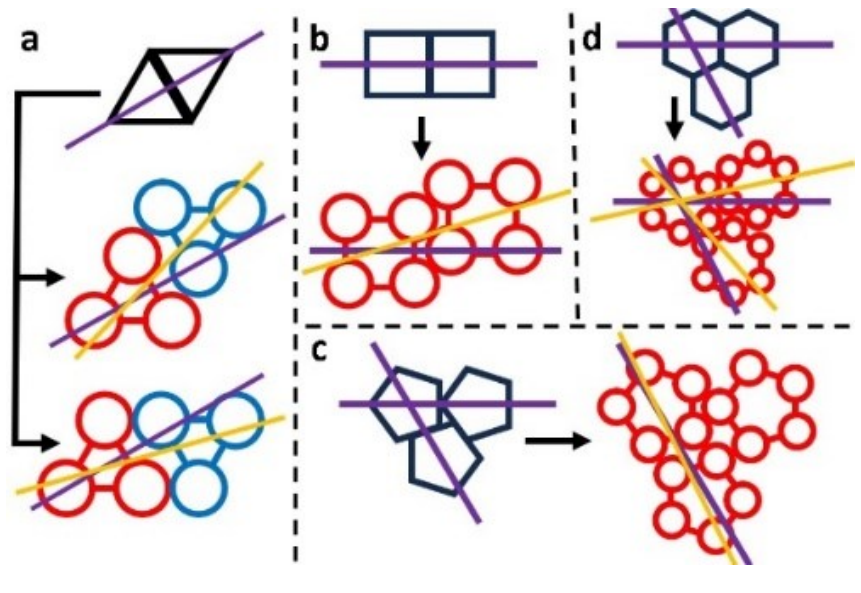

**Fig. 9. Schematic illustration of the confliction between shape and interaction.** **a** Triangle, in which two monomers coloured differently belong to the same unit-cell with different orientations. **b** Square. **c** Pentagon. **d** Hexagon. For each polygon, there is a non-zero torsional angle between the connecting line of neighbouring centre-of-masses determined by shape only (i.e., dense-packing pattern, represented by the dark purple line) and the one in the ground state of ball-stick polygon (represented by the orange line).

"More is different" *(55)*. People have accumulated extensive knowledge on both the entropy-limited case and the enthalpy-limited case for 2D systems, as well as the thermodynamic and spatial competitions between these two factors. Real 2D systems, however, stand in between, involving comparable entropy and enthalpy effects, whose phase behaviours are more abundant than either of the two above cases. This work reveals some interesting and unique physical features in 2D molecular systems, as well as establishes a theoretical framework which can be generalized to understand phase behaviours in more complex 2D systems. Besides its fundamental importance, our research is also valuable for the design of 2D materials at least in two aspects: (1) When there is a strong anisotropy induced by molecular shape or intermolecular interaction, the details of short-range interactions are crucial for designing various structures, such as amorphous materials used for frustrated magnetic materials *(45)*, and distorted lattice which is generally existed in some compound systems, e.g., high-entropy alloys *(56)*, widely used for tuning the mechanical properties *(57)* or the electron band structures *(58)*; (2) Phase stability at high pressures is mainly determined by shape, but at low pressures it can be tamed by the match of interaction and shape, which can facilitate the design of materials with desired physical features, such as ultra-stability and heat-sensitivity.

In our simulation, the deformation or elongation of the polygon can be ignored due to the large strengths of the bond and angle interactions (see Methods for more details). Since the mechanism we have provided above does not explicitly depend on bond length, we argue that our conclusion is still effective for the normal case of molecules. However, many studies have shown exotic melting scenarios and ground-state configurations when the polygon is allowed to deform *(59-61)*. Moreover, the melting scenarios of polygon mixtures may differ from identical polygon systems attributed to the much more complicated competition among various ingredients *(50)*. Considerations on these two factors may lead to a comprehensive understanding in the shape space and reveal the critical behaviour in the vicinity of $n_c$, but are beyond the scope of our study. In real applications, 2D materials have to be supported on a substrate. Although many substrates are known to have little influences on crystalline structures *(16, 20)*, some well-designed substrates may be used for taming melting scenarios *(62)*. The consideration on the effect of substrates might result in a more complex competition between entropy and enthalpy, leading to the emergence of new structures, and providing more guidance on experiments and industrial applications of 2D materials.

## Methods

### Preparation of Initial Configurations

Molecules are modelled by regular polygons with a 12-6 L-J particle on each vertex, and each intra-molecular pair of adjacent vertices are connected by a covalent bond. Each *n*-gon monomer is composed of *n* atoms and *n* covalent bonds. We use L-J units in this work, where physical quantities are expressed in a dimensionless manner, and set L-J parameters $m = 1$, $\sigma = 1$, and $\varepsilon = 5$. Both the covalent bonds and the valence angles are represented by a strong harmonic potential with $k = 1800\,\varepsilon/\sigma^2$. The equilibrium bond length in each polygon is set as short as $1.5\sigma$ to prevent the overlap of two monomers or two bonded atoms (A typical elongation of $\Delta l = 0.1\sigma$ would correspond to a characteristic temperature equal to $T = 18\varepsilon = 90$, significantly larger than the highest temperature we have used in all the simulations about $T$ = 20). We set the cutoff radius of the L-J interaction to $r_c = 7\sigma$ in the simulation, at which the L-J potential has decayed to a very low value (less than $10^{-4}$), ensuring a high accuracy on the estimation of potential energy. Moreover, we have performed several simulations with various values of $r_c$, from $3\sigma$ to $10\sigma$. A $r_c$ value out of this range is unnecessary to be tested because a smaller attraction range would result in a large error while a larger one should have no influences since the potential energy at $r_c = 10\sigma$ is as low as about $10^{-6}$. Packing modes of polygons with various $r_c$ values exhibit no qualitative differences, as shown in Fig. S16.

All the simulations started from the crystalline state unless otherwise specified and were performed at each temperature independently. To construct a crystalline structure, the unit-cell structure was first determined and then duplicated along two primary lattice vectors. In our polygonal systems, we determined the unit-cell structure by careful examination on the final structure from the REMD simulations described below. For triangle, the unit-cell contains two monomers with opposite body-orientations, as shown in Fig. 2c; for the other four polygons, the unit-cell is just a monomer. The exact values of the two primary lattice vectors were calculated by geometrical construction, taking the effective radius of each L-J particle into account. For the simulations of the triangular system starting from a crystalline structure with a locally disordered region, the initial configuration is prepared by the following procedure: At a relatively high temperature of $T$ = 2.88, the triangular system becomes amorphous spontaneously. When this happens, we select a snapshot with a small disordered region (about 30 monomers). This snapshot is then used as the initial configuration for related simulations.

**MD Simulations**

All the simulations in this work were performed with LAMMPS *(63)*. After the hand-made initial crystalline structures were optimized, they were tailored to fit in a simulation box with appropriate side-length ratio and tilting angle under periodic boundary condition. All the results reported in this paper were obtained from the simulations with 2496 polygonal molecules. Additional simulations on larger scales, 22500 monomers for the square system and about 5000 monomers for other polygons, found no qualitative differences in phase behaviours.

In *NPT* simulations, the shape of the simulation box controlled by the tilting angle and box sides ratio was invariant since the algorithm allows only isotropic shrinkage or stretching of the box *(63)*. Although the simulation box is perfectly commensurate with the ground-state crystal, there is no guarantee that it is still appropriate for crystalline morphology at relatively high temperatures. Therefore, other than *NPT* simulations for the triangular system, simulations in the isotension-isothermal ensemble *(43)* were performed for the other four systems. In each simulation, a very short *NVT* relaxation had been carried out for $6\times10^3$ steps before an *NPT* or isotension-isothermal equilibration run for $8.3\times10^6$ steps was performed. Since the typical correlation time was determined to be less than $1.6\times10^3$ steps, $2\times10^3$ configurations were then sampled with the interval of $2\times10^3$ steps. Because a longer simulation time might be required in each equilibration run, $4\times10^6$ more steps were carried out before sampling near the melting point. In very rare cases, the crystal melts at $T$ but stabilizes at a higher temperature $T'$. For these cases, we performed simulations at $T'$ initiating from the final configuration at $T$ or reseeded the random numbers used in the simulations. In this way, the system always melted/stabilized at the liquid state, which ensured that the previous simulation at $T'$ was stuck in the superheated state. This was also confirmed by the fact that the simulations at other temperatures slightly higher than $T$ resulted in melting as well. The Nosé-Hoover barostat and thermostat *[64-67]* were used to keep the system pressure and temperature constants. The simulation timestep was 0.005, much smaller than the characteristic time $\tau = 1/\sqrt{k}$, where $k$ is the strength of the valence bond, which is the fastest degree of freedom in the system.

For each polygon, the lowest pressure we study is $P = 0$, and the highest is $P = 10$ ($P = 8$ for octagon). An even higher pressure may cause numerical instability.

**REMD Simulations**

First, 625 monomers were put on the square lattice (25×25) with the same body-orientation, different from the equilibrium structure of any polygons to avoid the influence from the initial configuration. REMD simulations were then carried out in the *NVT* ensemble with 26 to 30 replicas. We have used uneven temperature intervals for different polygons to ensure that the exchange probabilities are all around 20%~35%, and the temperature distribution is shown in Fig. S1 in SI. The simulation temperature was kept constant by the Langevin dynamics. All the simulations lasted $7.2\times10^7$ steps and exchange attempts took place every $3\times10^5$ steps. For each polygon, after the simulation was finished, an energy minimization was performed on the final configuration at the lowest temperature to quench thermal fluctuations, and the optimized structures are shown in Fig. S2 in SI. Because the purpose of performing the REMD simulation was to investigate local structural features, the REMD simulation employed a reflect-wall boundary condition to avoid the influence of the box shape or the periodic boundary condition, which may prevent the formation of a dense but amorphous state like the spin-ice-like state that cannot fit into a regular box with periodic boundary condition seamlessly.

**Determinations of Phases and Phase Transitions**

The bond-orientational order parameter is defined as $\Psi_n = |< \frac{1}{N}\sum_i \psi_{n,i} >|$ and $\psi_{n,i} = \frac{1}{n}\sum_j e^{in\theta_{ij}}$, where $N$ is the total number of unit-cell, $i$ runs over all the representative points of the unit-cells while $j$ runs over the $n$ nearest neighbours of point $i$, < > means taking the ensemble average. For triangle, the representative points of the unit-cells are chosen as the COMs of molecules with the same initial body-orientation, and since we study the non-equilibrium dynamical process, $\Psi_4$ is calculated for each snapshot instead of taking the ensemble average; for the other four polygons, they are chosen as the COMs of all molecules. The characterizations of the spin-ice-like state of the triangular system and the distorted lattice of the square system have been described in the main text, and it is worth emphasizing that the Delaunay tessellation/Voronoi diagram in Fig. 5c is calculated by the freud library *(68)* with periodic boundary condition. Below we focus on the identifications of the crystal, hexatic, and liquid states. The liquid state has an approximately zero bond-orientational order parameter $\Psi_n$, the hexatic state has an exponentially decaying translational correlation function as well as a non-zero bond-orientational order parameter, and the crystal state has an algebraically decaying translational correlation function as well as a non-zero bond-orientational order parameter. The bond-orientational order parameter was calculated by the analysis tool in LAMMPS *(63)*.

The calculation of the translational correlation function is a key point for the phase determination of 2D melting *(69)*. In 2011, the pioneer work by Bernard and Krauth provided a method to evaluate the correlation function *(21)*, which was later modified by Li and Ciamarra for the L-J systems *(70)*. The correlation function is defined as $C_g(r,\theta) = <\delta(r-(x_i-x_j))\delta(r\tan\theta-(y_i-y_j))>$, where $\theta$ is the orientation of the crystal axis and should be determined by scanning $C_g(r,\theta)$ at a fixed $r$ (the first peak value corresponding to the orientation of the crystal axis).

Basically, a first-order phase transition can be identified by the discontinuous change in the caloric and density curves. We also confirm the discontinuous transitions by the corresponding order-parameters: body-orientational order parameter for solid-solid transition and bond-orientational order parameter for solid-liquid or hexatic-liquid transition.

The continuous transition for 2D crystal is theoretically a Kosterlitz-Thouless one instead of a strict second-order phase transition that can be evidenced by the divergence of heat capacity, so it is difficult to be identified by the potential energy or its derivative. Since we do not observe the KTHNY melting scenario in all our simulations, the only continuous phase transition we observe is the solid-hexatic transition. Therefore, we determine the phase transition point by the decaying behaviour of the translational correlation function: At the phase transition point, it suddenly changes from algebraical decay to exponential decay.

All the results for caloric curves as well as the calculations of order parameters and correlation functions can be found in SI (texts from S4 to S6 and figures from Fig. S10 to Fig. S15).

## Data Availability

MD trajectories corresponding to the determination of phase transitions and source data of all plots in both main text and SI have been deposited in Figshare under accession code https://doi.org/10.6084/m9.figshare.24460294. Source data are provided with this paper.

## Code Availability

The codes used in this study have been deposited in the Zenodo repository *(71)*.

**Acknowledgments**

We sincerely acknowledge Mr. Kun Tao, Prof. Fanlong Meng, Prof. Yuliang Jin, and Prof. Yilong Han for their helpful discussions and suggestions. The computations of this work were conducted on the HPC cluster of ITP-CAS and Tianhe-2 supercomputer. This work was supported by the National Natural Science Foundation of China (No. 11947302, Y.W.) and the Science and Technology Innovation Training Program of UCAS (R.Z.).

**Author contributions:**

Y.W. conceived and supervised the project. R.Z. performed simulations and analyzed the data. R.Z. and Y.W. discussed the results and wrote the paper.

**Competing interests:**

All authors declare that they have no competing interests.

**Supplementary Information**

**The PDF file includes:**
Supplementary Text
Sections S1to S6
Figs. S1 to S16

# Supplementary Information for

## A Critical Edge Number Revealed for Phase Stabilities of Two-Dimensional Ball-Stick Polygons

Ruijian Zhu[1,2] and Yanting Wang[1,2] *

[1] CAS Key Laboratory of Theoretical Physics, Institute of Theoretical Physics, Chinese Academy of Sciences, 55 East Zhongguancun Road, P. O. Box 2735, Beijing 100190, China
[2] School of Physical Sciences, University of Chinese Academy of Sciences, 19A Yuquan Road, Beijing 100049, China
* wangyt@itp.ac.cn

**Contents:**

**Supplementary Text**

## S1. Details and Results of REMD Simulations

The REMD simulations adopt uneven temperature intervals shown in Fig. S1, which are larger for higher temperatures, ensuring that the exchange probabilities are all within 20%–40%. The REMD simulations principally sample according to the Boltzmann distribution on each replica *(1)* , but the only purpose of performing them in this work is to generate the ground-state(s).

As reported in the main text, the final configuration at the lowest temperature after an energy minimization is very close to the ground state. The final structure of the triangular system (shown in Fig. 3a in the main text) is amorphous, which is a mixture of the two isotopic unit-cell structures. For square, the obtained structure is close to a square lattice with a tilted body-orientation. The monomers in the pentagon system are aligned into two body orientations to form alternative lines, and the spatial arrangement of their COMs is close to the triangular lattice. The hexagon system exhibits a triangular-lattice structure. In the obtained structure for the octagon, monomers basically arrange into a triangular lattice and share the same body orientation locally. The snapshots of the final structures and schematic illustrations of the local structures for square, pentagon, hexagon, and octagon, are shown in Fig. S2a-d, respectively.

## S2. Phase Behaviour of the Ball-Stick Triangular System

*Enumeration of the packing scheme*

As we have mentioned in the main text, there are three local packing schemes composed of the two isotopic unit-cell structures, which can further form three kinds of local structures: crystal, ‘perfect-mixing’, and ‘tower-like’. By enumeration, we demonstrate that: (1) The ‘perfect-mixing’ structure is not unique, instead, it contains a large number of different mixtures of the first two local structures shown in Fig. 2d in the main text, resulting in the ground-state degeneracy; (2) The third local structure in Fig. 2d cannot fill up the whole space by itself, it instead must mix with the first two. The enumeration rules can be summarized as below: (1) We initially choose one of the local structures in Fig. 2d, which is coloured in red in Fig. S3; (2) The ball-stick triangular molecules are added into the existing structure shell by shell without creating any geometrical conflicts, and each shell is coloured differently in Fig. S3. There are two possible conditions when adding monomers on the boundary, as shown in Fig. S3a. The first one is deterministic: If we place the orange molecule on the boundary, then as guided by the orange line, the positions of the three brown molecules are completely determined by the existing monomers. The second one is non-deterministic: If we add the blue molecule on the boundary, as guided by the blue line, we have two different ways to add the three purple molecules on the boundary without creating any geometrical conflictions, as labelled by 1 and 2 in Fig. S3a. Here “deterministic” or “determine” means that there is only one possible position for the newly added molecule if the creation of geometrical defects is disallowed.

The three structures illustrated in Fig. S3b-d are all instances of the ‘perfect-mixing’ structure. Starting from the two adjacent third local structures in Fig. 2d, we add monomers one by one as illustrated in Fig. S3e: The three monomers connected by the green line determine the position of the next monomer shadowed in green, and then the yellow line determine the position of the next monomer shadowed in yellow. After that, we determine the positions of the monomers coloured in purple and brown in sequence. The boundary structure circled in blue has a geometrical confliction, which is inevitable since every step in this process is deterministic. Therefore, the third local structure in Fig. 2d is not space-filling. Moreover, by mixing the three local structures in Fig. 2d, we create a large ‘tower-like’ pattern shown in Fig. S3f, which is deterministic up to at least the fourth shell, so there has very likely only one ‘tower-like’ structure if we ignore the chirality degree of freedom.

*Results of regular MD simulation at $P$ = 10*

The regular MD simulation for the ball-stick triangular system is also performed under $P = 10$ to test the robustness of the observed phase behaviour. As shown in Fig. S6a, the caloric and density curves both suggest a discontinuous phase transition at $T = 5.54$. The amorphization occurs at lower temperatures, leading to a disordered spatial arrangement of monomer COMs, analogous to the case for $P = 0$. This phase transition is also evidenced by the sudden drop of $\Psi_{6,\mathrm{atom}}$, as plotted in Fig. S6b.

**S3. Phase Behaviour of the Ball-Stick Square System**

*Additional results for P = 0*

The simulation on a system with $N = 22500$ gives similar results with $N = 2496$. As shown in Fig. S8a, the caloric curve, density curve, and bond-orientational order parameter all show discontinuous changes at $T = 2.89$, demonstrating a first-order phase transition. The concentration of topological defects is calculated at various temperatures, as shown in Fig. S8b, providing values higher than 0.4. It is important to notice that, although the Voronoi cell has a less dispersive area distribution at lower temperature, as shown in Fig. S8c, the concentration of topological defects is higher and reaches about 0.5 at $T = 0.01$ with the simulated annealing procedure, whose representative snapshot is shown in Fig. S8d, suggesting no visible differences from a square lattice. We also perform simulations at $T = 2.5$ without periodic boundary condition for these two sizes, whose results are also plotted in Fig. S8b. The perfect agreement among different sizes and different boundary conditions manifests that the distortion is caused by the intrinsic property of the system rather than the system size or the boundary condition. This is also the case for the translational correlation function, where the curves obtained from $N = 2496$ and $N = 22500$ at the same temperature coincide with each other exactly, as plotted in Fig. S8e.

Finally, we perform a ‘renormalization’ process for the translational correlation at $T = 2.8$, which is relatively far from the phase transition point, sharing the same procedure described in the main text. The result in Fig. S8f clearly demonstrates self-similarity after rescaling the distance with the characteristic length of the coarse-grained system, which indicates that the square system locates as a critical point in the shape space rather than the thermal-parameter space.

*Simulation results for P = 5 and P = 10*

We perform simulations beginning from the crystalline morphology at $P = 5$ and $P = 10$ to test the robustness of the phase behaviour and the distorted lattice structure. As shown in Fig. S9a, both the caloric and density curves at $P = 5$ and $P = 10$ indicate discontinuous phase transitions, whose melting points are 4.55 and 6.16, respectively. These two discontinuous changes are in agreement with the sharp drop of the bond-orientational order parameter at corresponding pressures, as shown in Fig. S9b. The translational correlation function $C_g(r)$ plotted in Fig. S9c behaves similarly as $P = 0$. The calculated concentration of topological defects shows little differences at various pressures near the melting point, as shown in Fig. S9d.

## S4. Phase Behaviour of the Ball-Stick Pentagon System

As we have described in the main text, the phase behaviour of the ball-stick pentagon system is the same as the hard pentagon *(2, 3)*, in which a striped phase transforms into a rotator crystal phase before melting into liquid. At each pressure, we identify two discontinuous changes in the caloric curves plotted in Fig. S10a (from the striped phase to the rotator crystal phase) and Fig. S10b (from rotator crystal to liquid), corresponding to the two first-order transitions. The first phase transition can be evidenced by the body-orientational order parameter (plotted in Fig. S10c) defined as $\Phi_n \equiv |< \frac{1}{N}\sum_i \phi_i >|$ where $\phi_i \equiv \exp(in\theta_i)$, $n$ corresponds to the rotational symmetry of monomer (we use $n = 10$ for the striped phase because there have a five-fold symmetry and two body-orientations) and $\theta_i$ is the angle between $x$-axis and a vector from the monomer COM to an arbitrary vertex. The phase transition points are $T_{m1} = 2.39$ at $P = 0$, $T_{m1} = 3.41$ at $P = 5$, and $T_{m1} = 5.33$ at $P = 10$. The second phase transition is quantified by the bond-orientational order parameter, as shown in Fig. S10d, where the phase transition points are $T_{m2} = 3.9$ at $P = 0$, $T_{m2} = 8.01$ at $P = 5$, and $T_{m2} = 12.58$ at $P = 10$. At $P = 10$, the system switches between the rotator crystal state and the liquid state at $T = 12.58$, reflecting the coexistence at the first-order phase transition temperature. As a result, it is difficult to decompose the bond-orientational order parameter of each snapshot into two states, so we do not plot it in Fig. S10d.

At $P = 5$ and $P = 10$, the distribution of the body-orientation at each pressure still shows two peaks in the striped phase, while it is totally random in the rotator crystal phase, as shown in Fig. S10e, analogous to the case at $P = 0$. The translational correlation functions in these two states plotted in Fig. S10f-h decay algebraically, ensuring that there is no hexatic phase.

## S5. Phase Behaviour of the Ball-Stick Hexagon System

*Low pressures* ($P = 0$–$1$)

At relatively low pressures, as a common feature for the L-J system *(4, 5)*, the hexatic phase does not appear. Based on the caloric curves in Fig. S11a, we identify a sharp transition at each pressure: $T_m = 5.36$ at $P = 0$, $T_m = 5.66$ at $P = 0.5$, and $T_m = 5.96$ at $P = 1$. The bond-orientational order parameter in Fig. S11b also shows a discontinuous change at $T_m$ for each pressure, in agreement with the first-order nature and suggesting that the system is in the fluid state when $T > T_m$. All the translational correlation functions in Fig. S11c exhibit the algebraically decaying behaviour, demonstrating that when $T < T_{\mathrm{m}}$, the system is in the crystal state. Moreover, it is worth noticing that, at $P = 0$ and $T > T_{\mathrm{m}}$, the system has an approximately zero potential energy. We further calculate the RDFs of the system after melting at $P = 0$ and $P = 0.5$. As shown in Fig. S11d, the curve at $P = 0$ has only one peak, corresponding to the gas state; while at $P = 0.5$, the curve has several peaks, corresponding to the liquid state. These features indicate that the crystal directly sublimates at $P = 0$, and the liquid state only appears at $P > 0$.

*Intermediate pressures* ($P = 1.5$–$5$)

At intermediate pressures, a rotator crystal phase arises, which is observed in both the simulation and experiment of the round-corner hexagon *(6)* but does not appear in the simulation of hard hexagon *(2)*. As shown in Fig. S12a, the triangular lattice crystal experiences a first-order phase transition into the rotator crystal phase ($T_{\mathrm{m1}} = 6.27$ at $P = 1.5$, $T_{\mathrm{m1}} = 6.89$ at $P = 2.5$, and $T_{\mathrm{m1}} = 8.42$ at $P = 5$). In the triangular lattice crystal state, the monomers have only one body-orientation, as shown in Fig. S12b, and the fluctuation is smaller at lower temperatures and pressures. In the rotator crystal phase, the monomers have two possible body-orientations, as shown in Fig. S12c, and it is interesting that these two peaks are more obvious at a higher pressure and temperature. Moreover, the bimodal distribution instead of a totally random one leads to a non-zero body-orientation order parameter $\Phi_6$ even at a temperature higher than the phase transition point.

Further heating up the rotator crystal, it will melt into liquid discontinuously, as shown in the caloric curve plotted in Fig. S12d ($T_{\mathrm{m2}} = 6.4$ at $P = 1.5$, $T_{\mathrm{m2}} = 7.72$ at $P = 2.5$, and $T_{\mathrm{m2}} = 11.31$ at $P = 5$). At $P = 5$, we do observe the switch between two states respectively at $T = 11.3$, 11.31, and 11.32, as an important feature for phase coexistence in *NPT* simulations. The solid-liquid phase transition is evidenced by the bond-orientational order parameter plotted in Fig. S12e, and the translational order also vanishes at the same temperature, as shown in Fig. S12f.

*High pressures* ($P = 7.5$–$10$)

At high pressures, it takes three steps to melt from triangular lattice to liquid. At $T_{\mathrm{m1}}$ (9.94 at $P = 7.5$ and 11.42 at $P = 10$), a solid-solid phase transition occurs, in which the triangular lattice transforms to the rotator crystal phase via a discontinuous phase transition. This transition shows no qualitative differences from the case at lower pressures. At $T_{\mathrm{m2}}$ (14.69 at $P = 7.5$ and 18.14 at $P = 10$), the rotator crystal melts into the hexatic state continuously, and the melting point can be identified by the decaying behaviour of the translational function, which is algebraical in the crystal state while exponential in the hexatic state, as shown in Fig. S13a. At $T_{\mathrm{m3}}$ (14.99 for $P = 7.5$ and 18.55 for $P = 10$), the hexatic state melts into the liquid state discontinuously, and the corresponding caloric curves are shown in Fig. S13b. It is obvious that the discontinuous nature is weaker at a higher pressure, but the frequent switches between the two states at the transition points indicate that, even at $P = 10$, the transition is still a first-order one instead of a KT one.

## **S6. Phase Behaviour of the Ball-Stick Octagon**

*Low pressures* ($P$ = 0–2.5)

The triangular lattice crystal transforms into a rotator crystal phase at $T_{m1}$ (2.71 at $P$ = 0, 3.02 at $P$ = 1, and 3.46 at $P$ = 2.5) as evidenced by the caloric curve and the body-orientational order parameter $\Phi_8$, as plotted in Fig. S14 a and b, respectively. As shown in Fig. S14c, the caloric curve suggests that the rotator crystal melts into liquid directly at $T_{m2}$ (3.83 at $P$ = 0, 5.69 at $P$ = 1, and 8.91 at $P$ = 2.5), which is also manifested by the bond-orientational order parameter shown in Fig. S14d. The quasi-long-range translational order retains until the melting point, as shown in Fig. S14e, excluding the possibility of the appearance of intermediate phases. Moreover, it is worth mentioning that, as shown in Fig. S14f, the distribution of the body-orientation is very rough and shows no evidence of the three peaks at higher temperatures and pressures. Whether it is intrinsic or caused by the finite-size effect is unclear and left for future studies.

*High pressures* ($P$ = 5–8)

The L–J disk system follows a solid-liquid transition at low temperatures (pressures), while it follows a hard-disk-like behaviour at high temperatures (pressures) *(4, 5)*. This is also the case in the ball-stick octagon system. During heating, the triangular lattice crystal first transforms into the rotator crystal at $T$ = 4.16 for $P$ = 5 and $T$ = 4.92 for $P$ = 8. Then the rotator crystal loses its translational order at $T$ = 14.54 for $P$ = 5 or $T$ = 20.99 for $P$ = 8 to transform into the hexatic phase, and finally loses its bond-orientational order at $T$ = 14.55 for $P$ = 5 or $T$ = 21.27 for $P$ = 8 to transform into the liquid state. The solid-solid transition is similar to the cases at low pressures, as shown in Fig. S15a. The transition from rotator crystal to hexatic is a KT one, whose transition point is determined by the change of the decaying behaviour of the translational correlation function shown in Fig. S15b. The transition from hexatic to liquid is a discontinuous one, indicated by the sudden change in the caloric curve plotted in Fig. S15c.

**Figures**

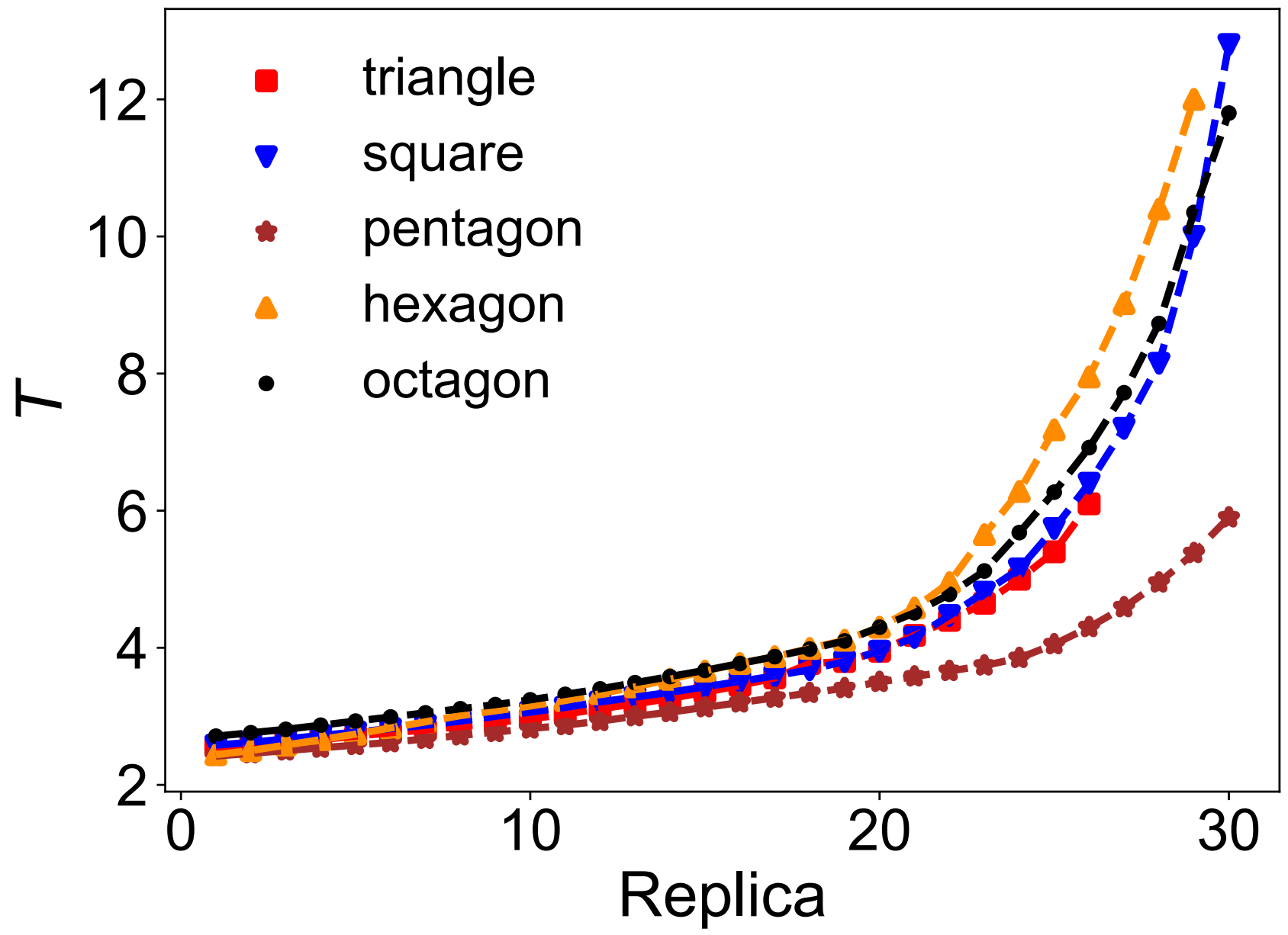

**Fig. S1. Temperature distributions in the REMD simulations.**

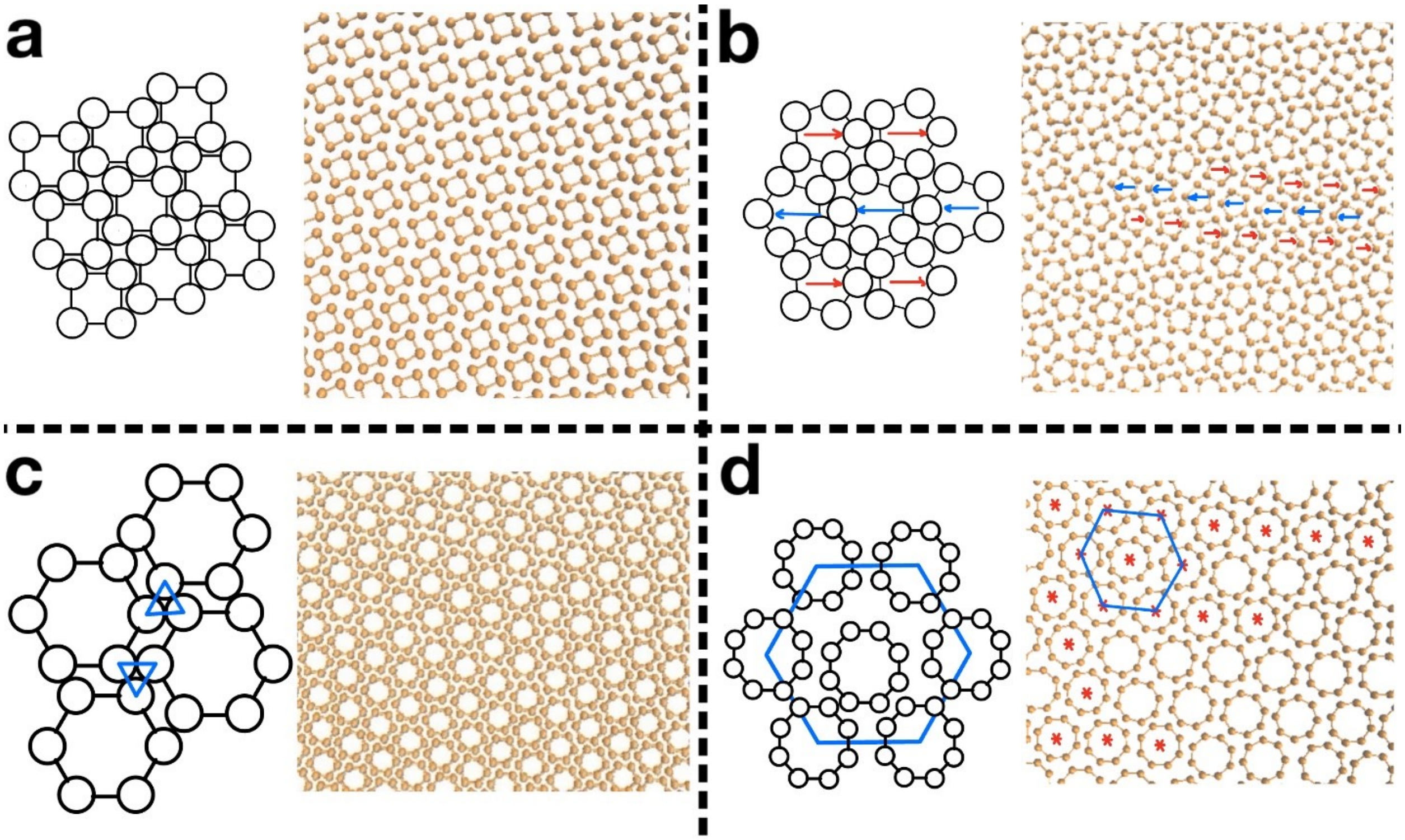

**Fig. S2. Typical snapshots of the stable structures obtained from the REMD simulations.** In each figure, the left panel is a schematic illustration of the local packing scheme, while the right one is a part of the snapshot. **a** Square. **b** Pentagon, in which we also label the body-orientation of a small part of the monomers to see clearer. **c** Hexagon, in which we label the local atom cluster. **d** Octagon, in which we also label the particles sharing the same body-orientation locally by red crosses and the hexagonal distribution of the six-nearest neighbours by the blue hexagon.

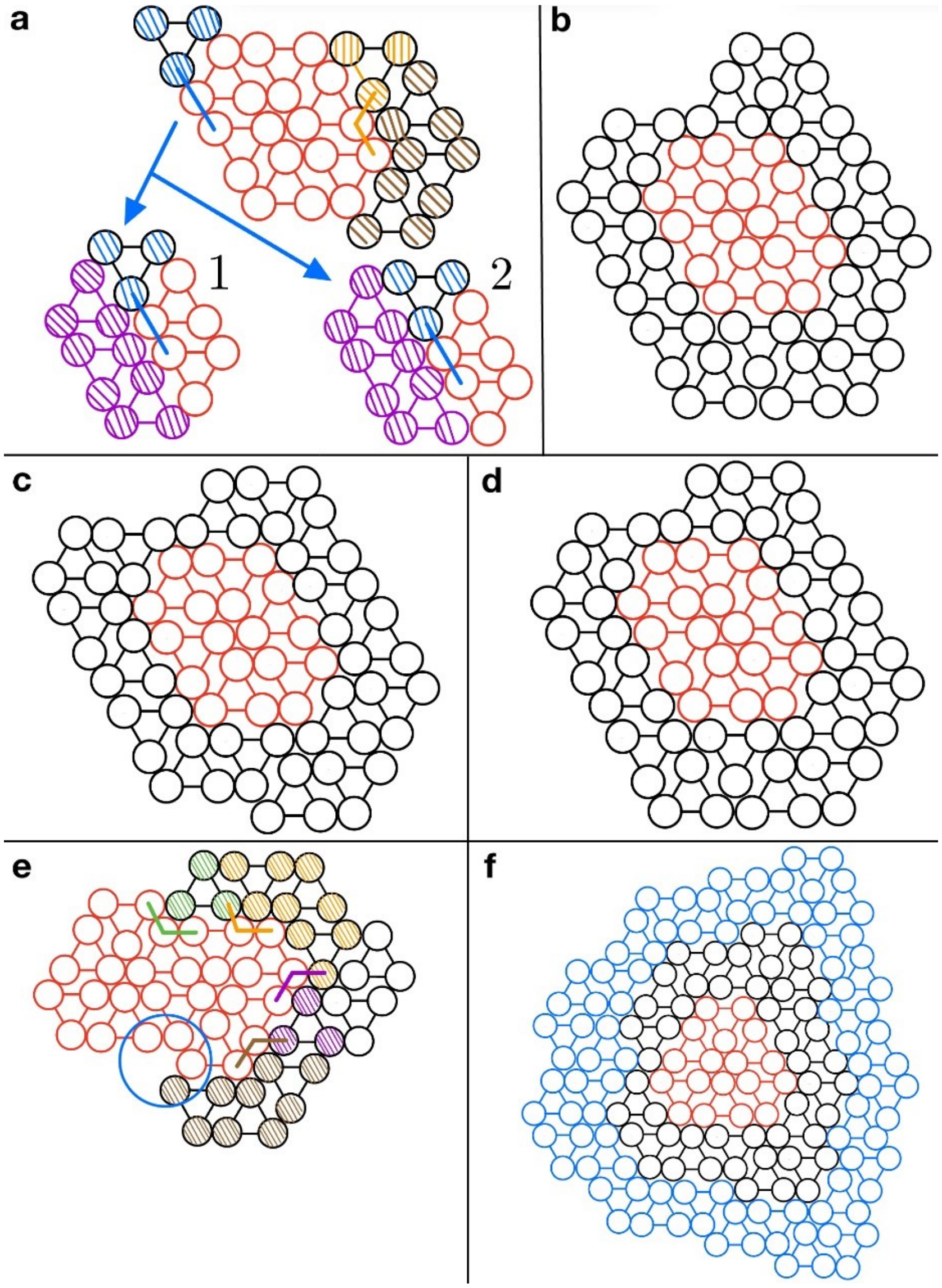

**Fig. S3. Enumeration of different combinations of the three local structures in the triangular system. a** Illustration of the deterministic (orange molecule, orange line, and brown molecules) and non-deterministic (blue molecule, blue line, and purple molecules) local structures on the boundary. The non-deterministic local structure can lead to two different schemes when adding molecules around it, labelled by 1 and 2. **b-d** are formed only by the first two local structures, noted as 'perfect-mixing', while **f** is formed by all the three local structures, noted as 'tower-like'. In each figure, the red monomers are initially fixed, and more monomers are added around them shell by shell. The monomers in the second shell are coloured black and those in the third shell are coloured blue, as shown in **b-d** and **f**. **e** Enumeration to show that the third structure in Fig. 2d in the main text cannot fill up the whole space by itself. Each line "determines" the positions of molecules with the same colour, and the blue circle highlights the unavoidable geometrical defect. We also emphasize that **b-d** are just some examples for the 'perfect-mixing' structure. In fact, there have more different structures even if we only enumerate to the second shell.

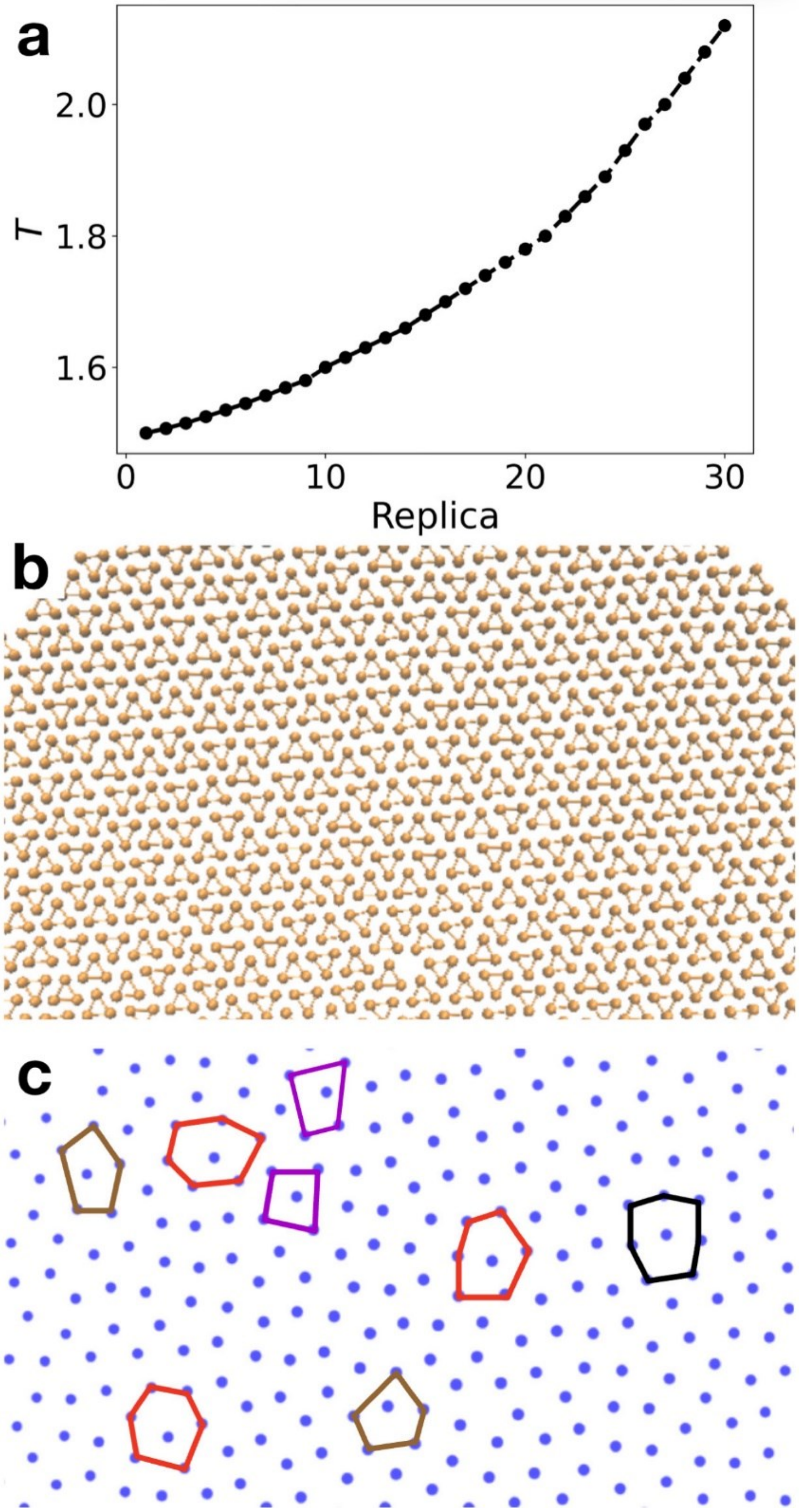

**Fig. S4 The results of the REMD simulation for the triangular system at much lower temperatures.** **a** Temperature distribution in the REMD simulation. **b** A typical snapshot from the replica with the lowest temperature ($T$ = 1.5), where the molecules arrange densely but the two isotopes mix randomly. **c** The COMs of the triangular system, exhibiting a highly random arrangement.

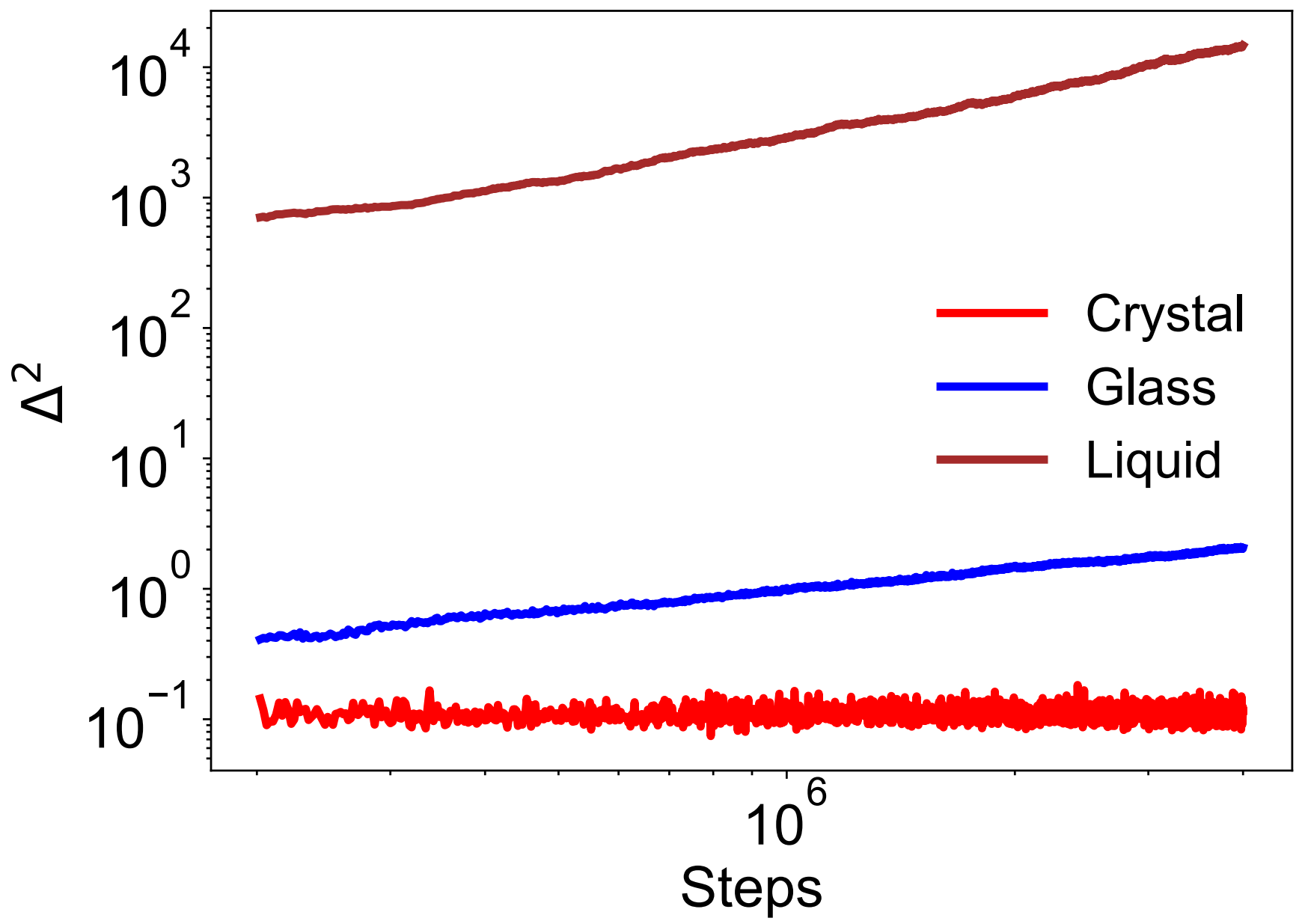

**Fig. S5 MSDs for the triangular system in different states.** The large gaps between different lines clearly distinguish three different dynamical behaviours: random diffusion in the liquid state, frustrated diffusion in the glassy-like state, and local rattling in the crystal state.

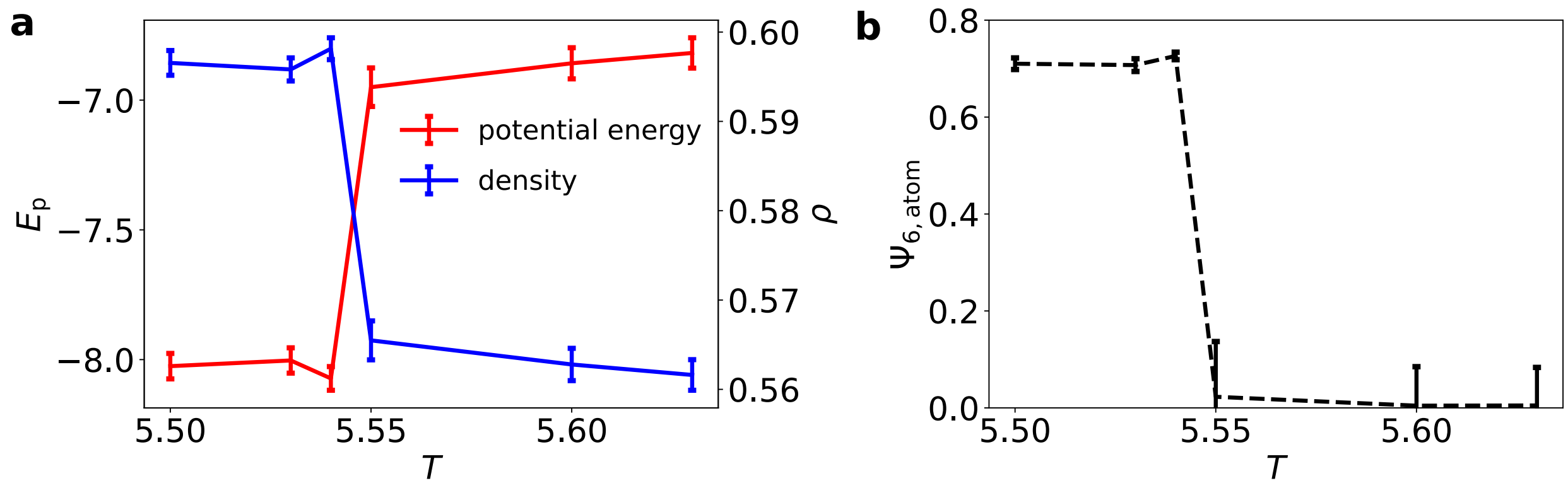

**Fig. S6. Phase behaviour of the ball-stick triangular system at *P* = 10. a** Caloric curve and density curve. **b** Bond-orientational order parameter defined for all atoms vs. temperature. The discontinuous changes all are located at $T = 5.54$–$5.55$.

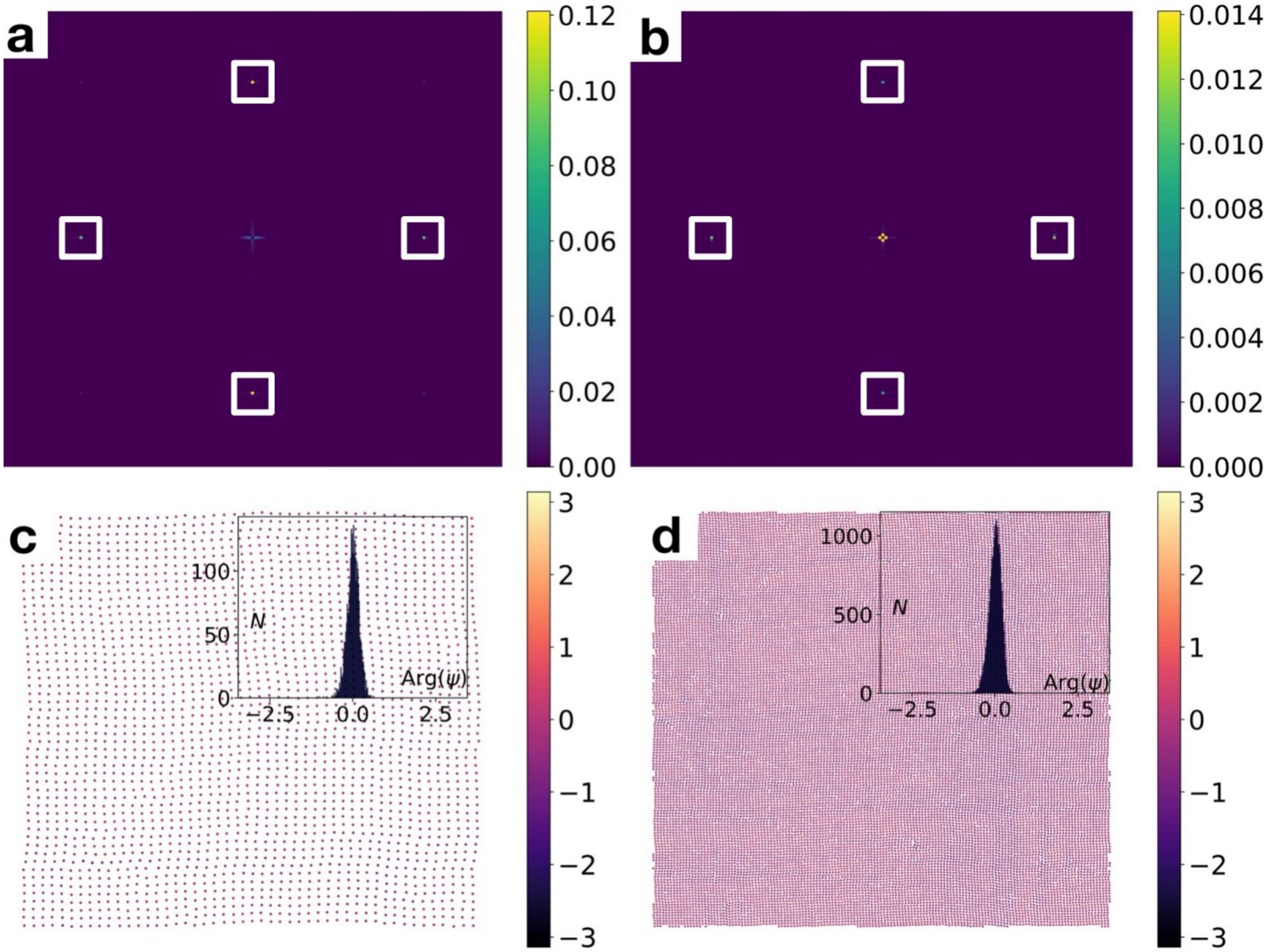

**Fig. S7. Four-fold symmetry of the square system.** **a** and **b** are the diffraction patterns, in which the white squares highlight the light spots. **c** and **d** are the heatmaps coloured according to the arguments of bond-orientational order parameter of monomers, whose distributions are plotted in the insets. **a** and **c** for $N = 2496$, **b** and **d** for $N = 22500$.

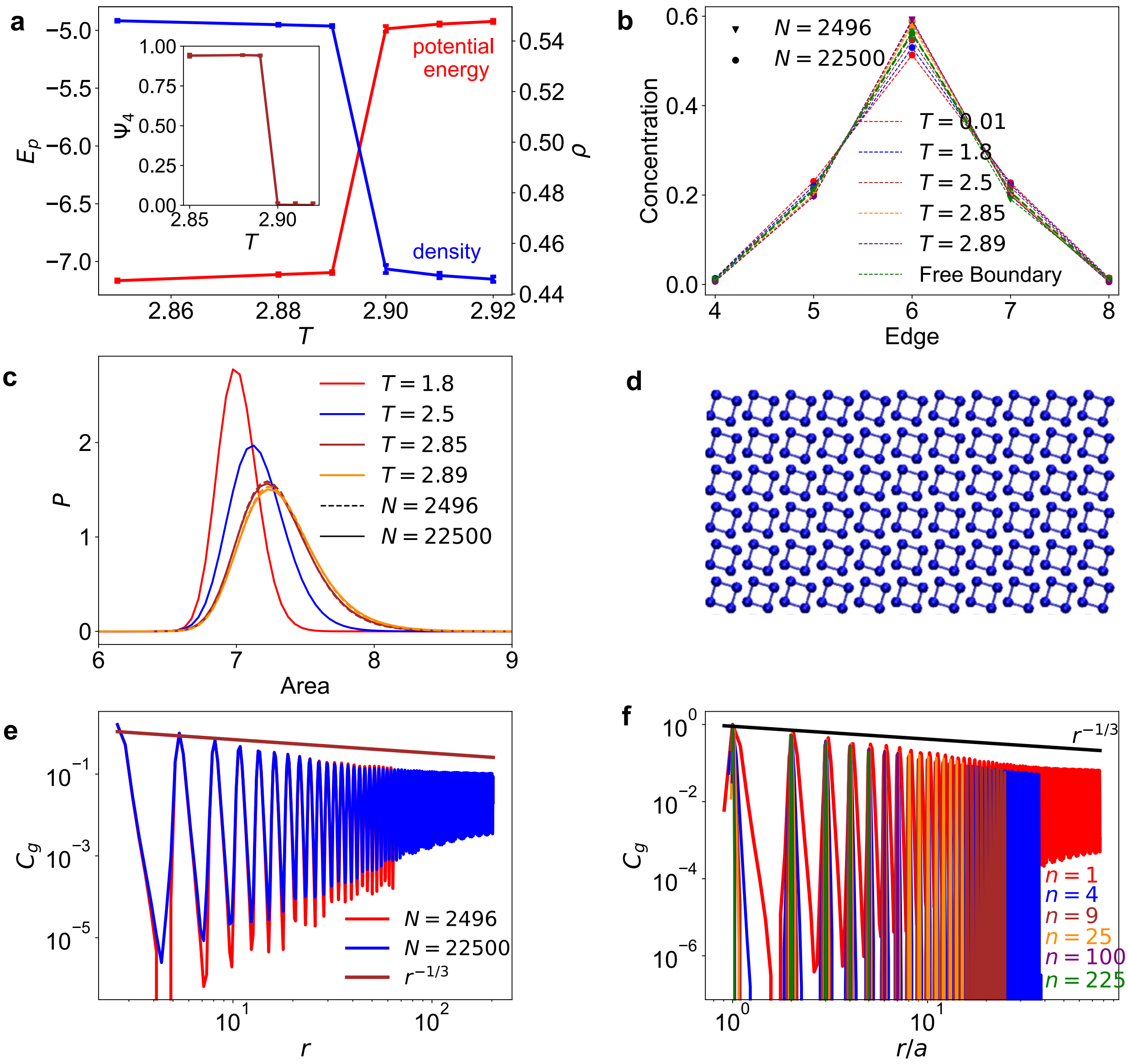

**Fig. S8. Additional simulation results for the square system at $P$ = 0. a** Caloric curve and density curve for $N$ = 22500. The inset is the bond-orientational order parameter at different temperatures. All the discontinuous changes of these three curves locate at $T = 2.89 \sim 2.9$. **b** The probability density of the edges of the Voronoi cell at various temperatures, simulation scales, and boundary conditions. **c** The distribution of the area of the Voronoi cell at various temperatures and simulation scales. **d** A typical snapshot of the simulated annealing simulation at $T = 0.01$. **e** The translational correlation function of two different system sizes at the same temperature of $T = 2.89$. **f** The translational correlation function of $N = 22500$ at $T = 2.8$ with different block sizes.

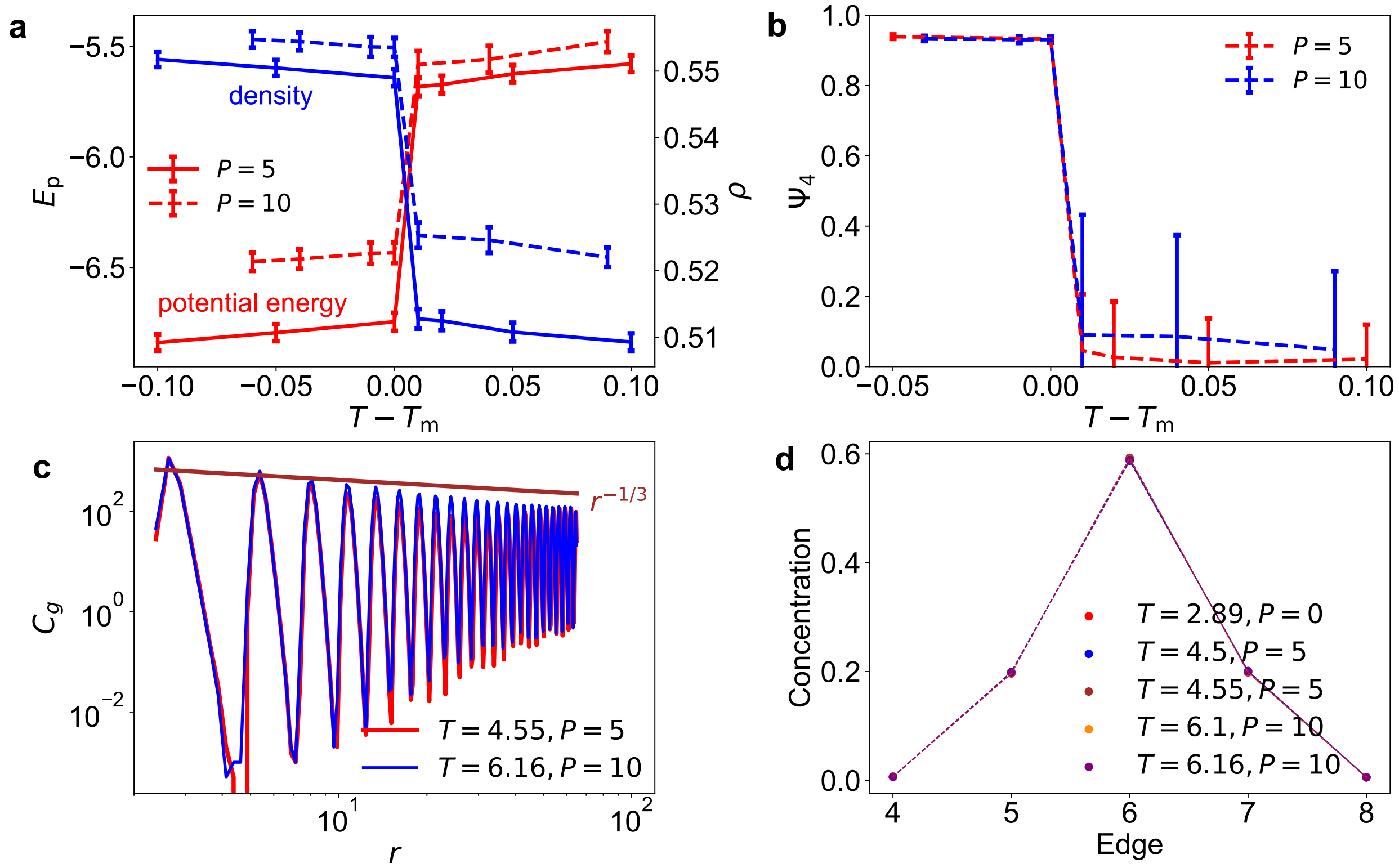

**Fig. S9. Phase behaviour of the ball-stick square system at *P* = 5 and *P* = 10. a** Caloric curve and density curve. **b** Bond-orientational order parameter vs. temperature at $P = 5$ and $P = 10$. **c** Translational correlation function, whose decaying behaviour is generally same as the case for $P = 0$. **d** The probability density of the edges of the Voronoi cell at various pressures and temperatures near the melting point.

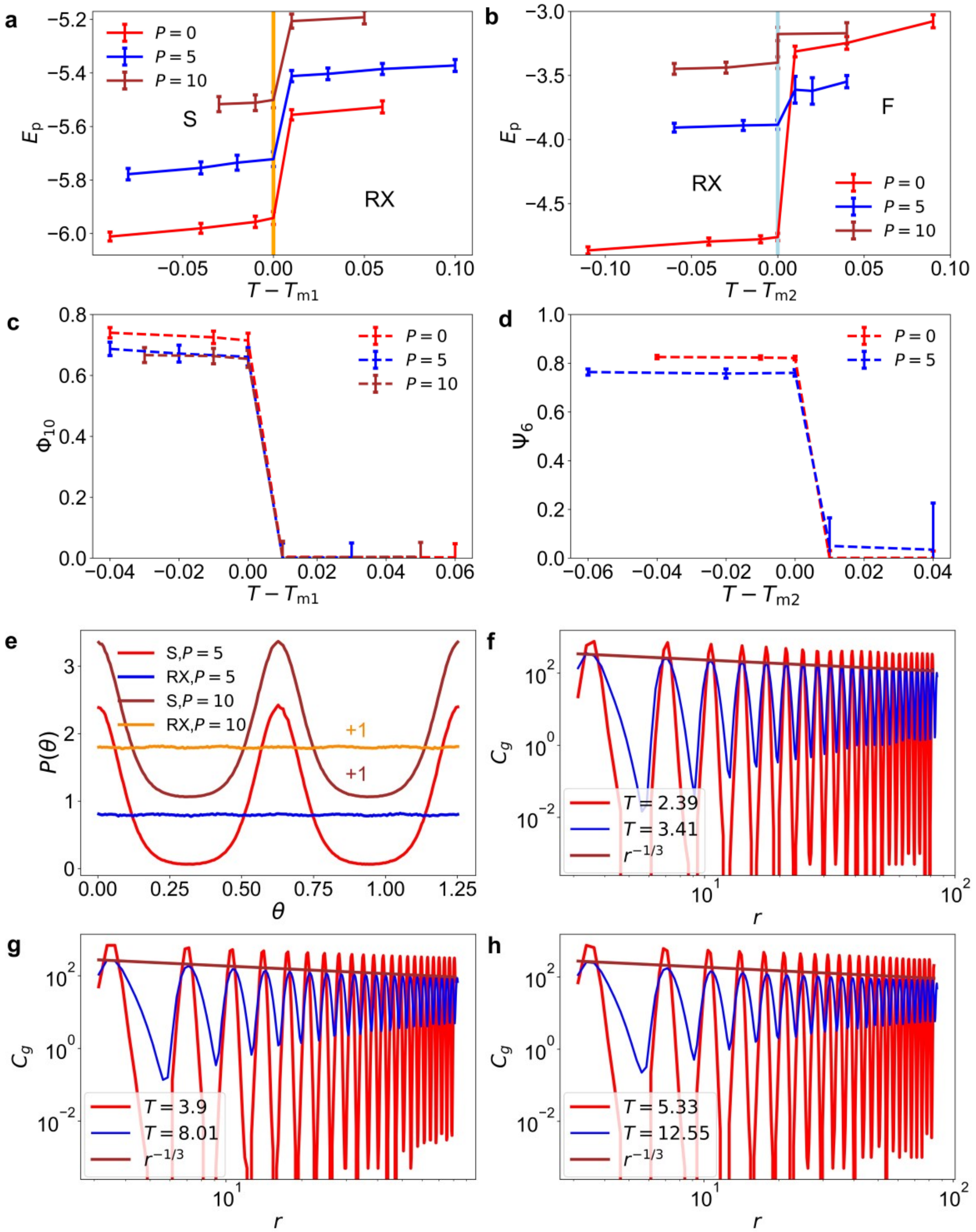

**Fig. S10. Phase behaviour of the ball-stick pentagon system.** **a** Caloric curves under different pressures, manifesting discontinuous transitions from the striped phase to the rotator crystal phase. **b** Caloric curves under different pressures, manifesting discontinuous transitions from the rotator crystal phase to the liquid phase. **c** Body-orientational order parameter characterizing the phase transition from the striped phase to the rotator crystal phase. **d** Bond-orientational order parameter characterizing the phase transition from the rotator crystal phase to the liquid phase. **e** Distribution of the body-orientation at $P = 5$ and $P = 10$ in different solid states. **f**–**h** Translational correlation functions at different solid states at $P = 0$, $P = 5$, and $P = 10$, respectively, all of which show algebraically decaying behaviour manifesting a solid state instead of a hexatic one.

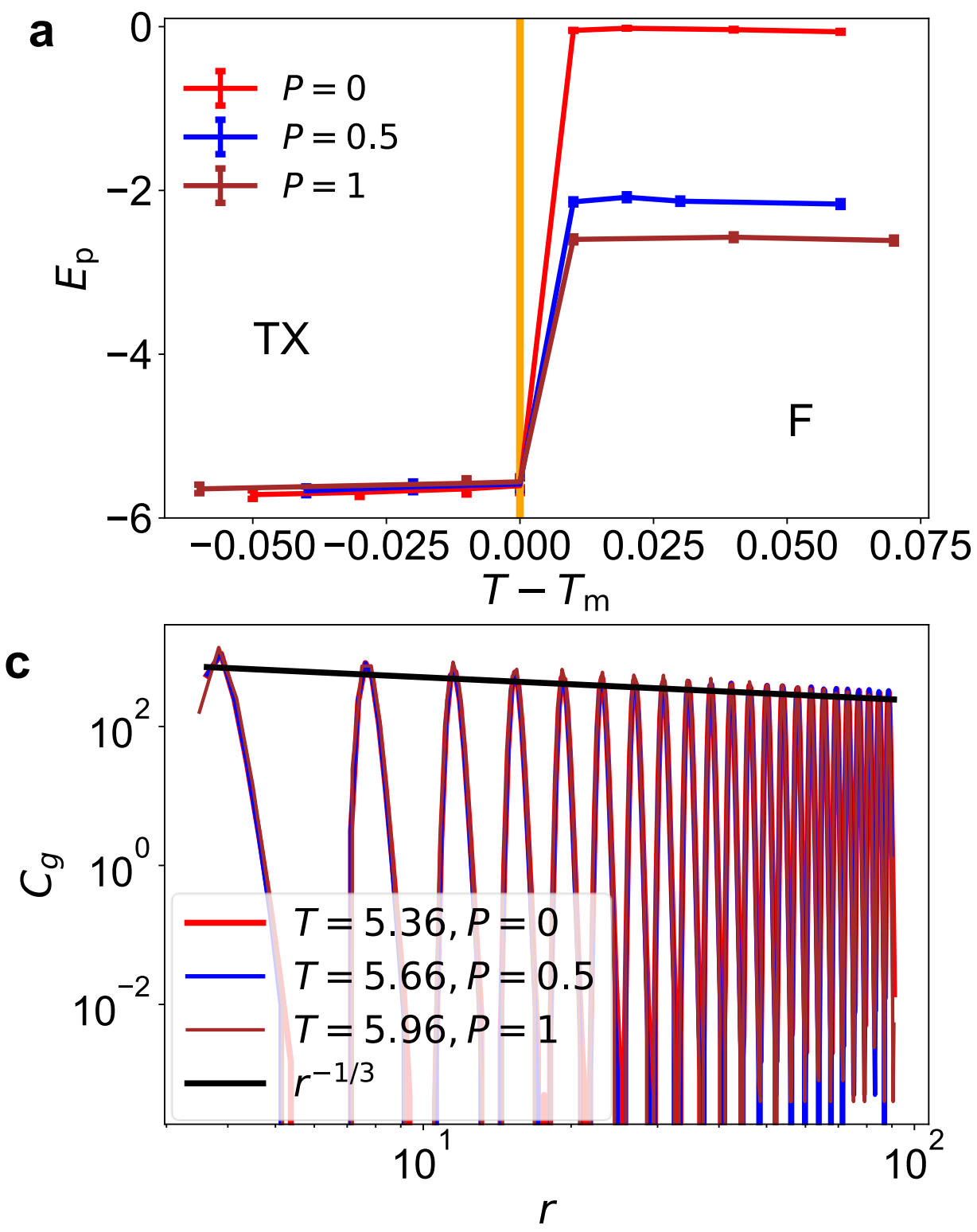

**Fig. S11. Phase behaviour of the ball-stick hexagon system at low pressures. a** Caloric curves at different pressures, showing the phase transitions from triangular lattice crystal to fluid. **b** Bond-orientational order parameter vs. temperature. **c** Translational correlation functions. **d** RDFs of gas ($P = 0$) and liquid ($P = 0.5$).

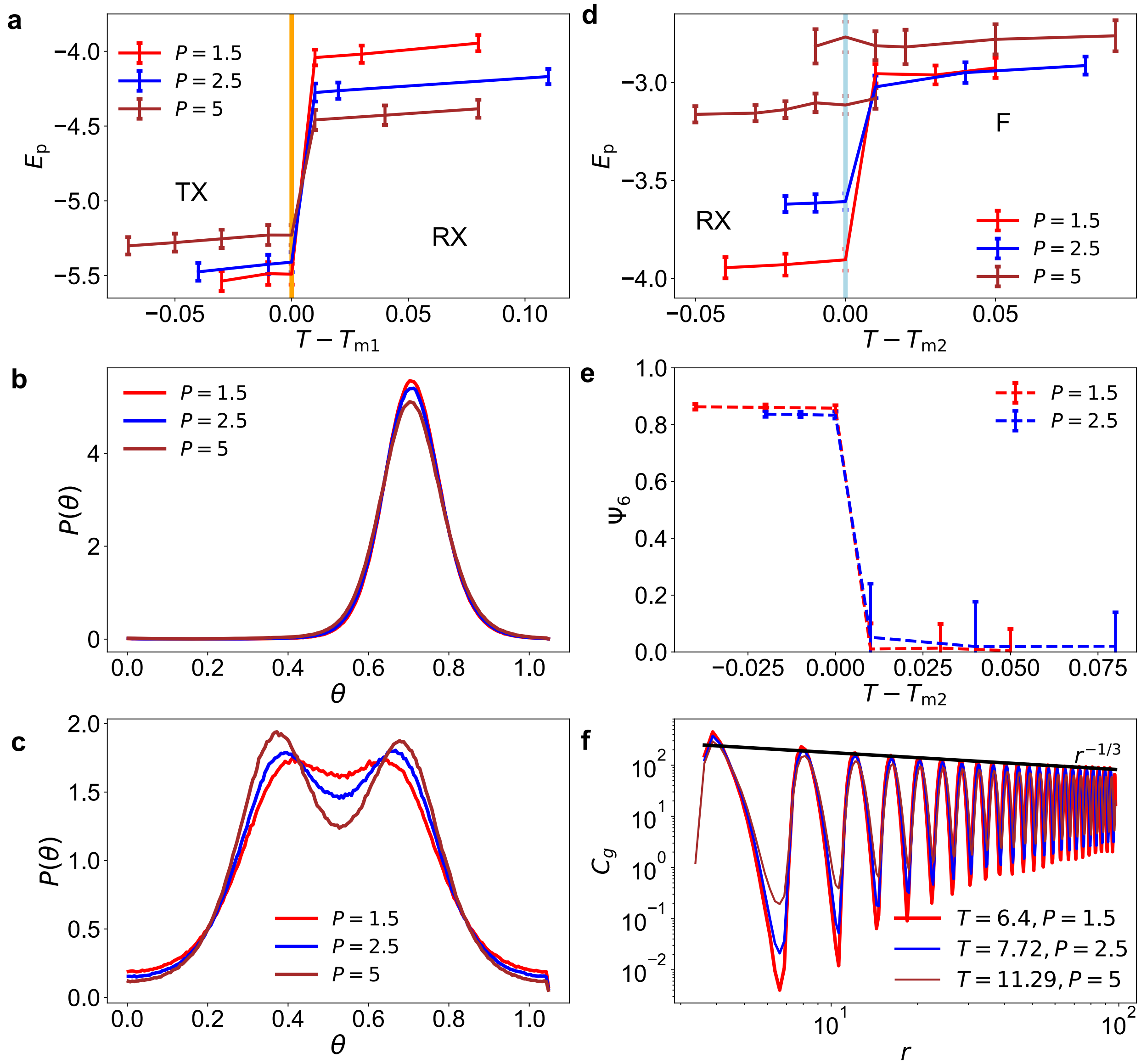

**Fig. S12. Phase behaviour of the ball-stick hexagon system at intermediate pressures. a–c** Triangular lattice crystal to rotator crystal. **d–f** Rotator crystal to liquid. **a** and **d** Caloric curve. **b** and **c** Distribution of the body-orientation in the triangular lattice crystal phase and the rotator crystal phase, respectively. **e** Bond-orientational order parameter vs. temperature. **f** Translational correlation functions in the rotator crystal phase.

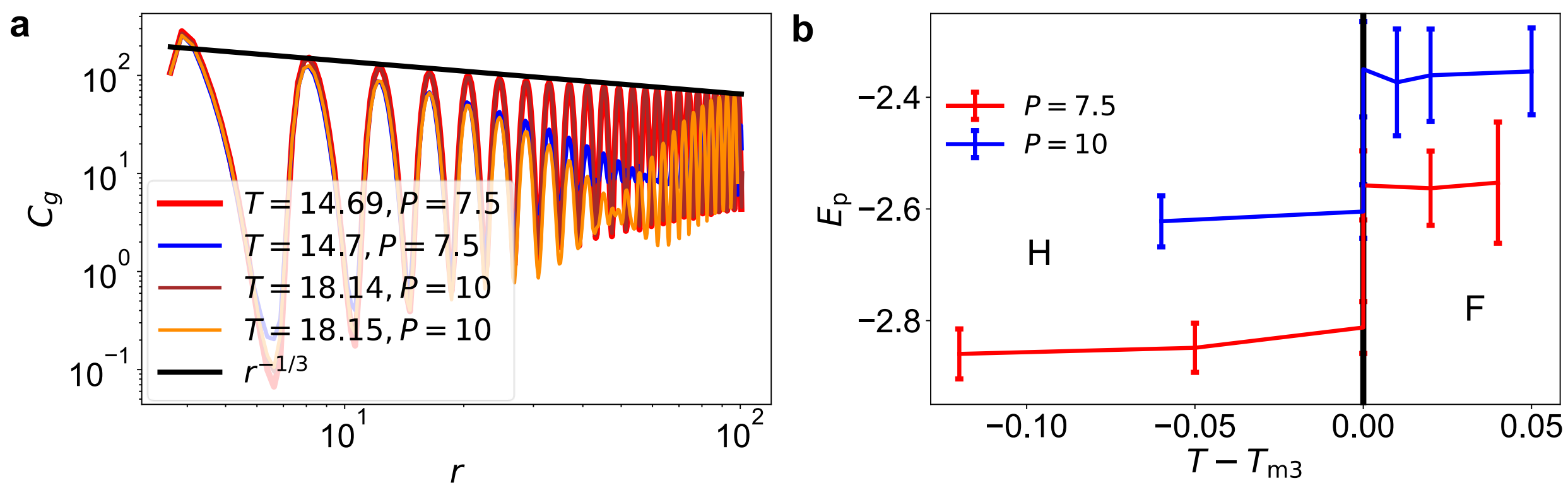

**Fig. S13. Phase behaviour of the ball-stick hexagon system at high pressures.** We use the translational function **(a)** to identify the solid-hexatic phase transition (algebraically decay vs. exponentially decay) and the caloric curve **(b)** to characterize the hexatic-liquid transition.

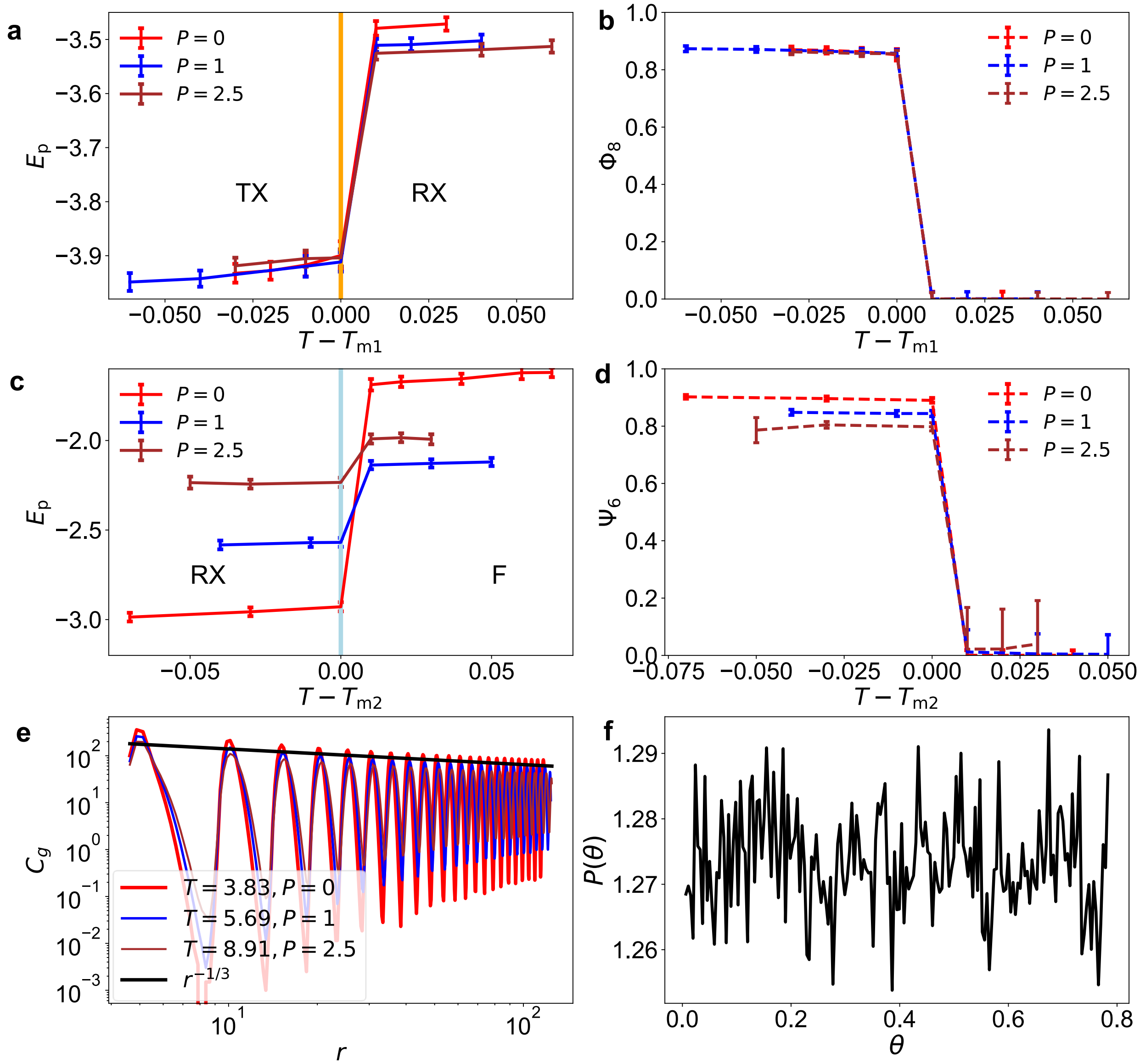

**Fig. S14. Phase behaviour of the ball-stick octagon system at low pressures. a** Caloric curves at different pressures for the solid-solid phase transition. **b** Body-orientational order parameter vs. temperature. **c** Caloric curves at different pressures for the melting of rotator crystal. **d** Bond-orientational order parameter vs. temperature. **e** Translational correlation functions. **f** Distribution of the body-orientation at $P = 2.5$ and $T = T_m$ (8.91).

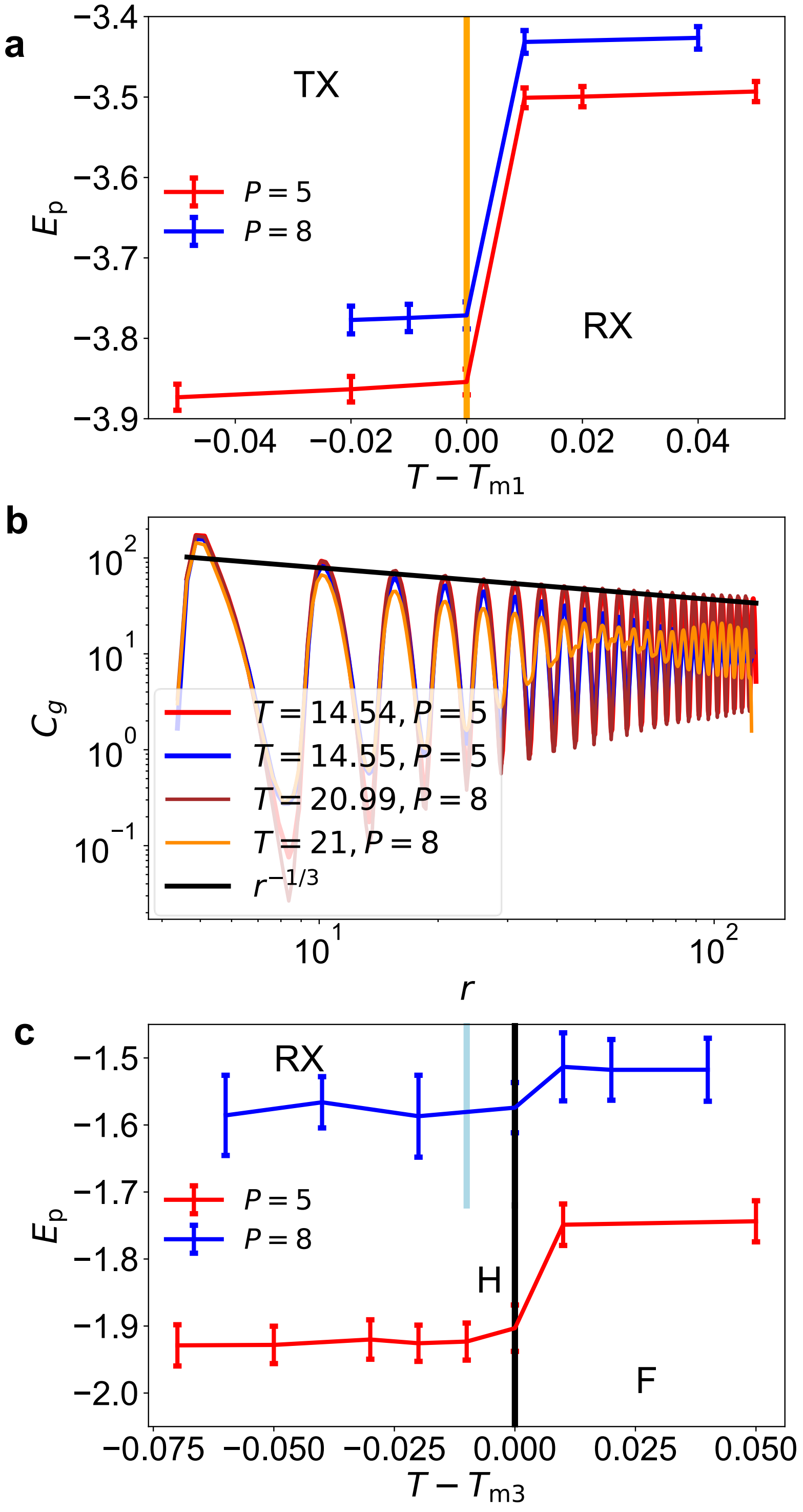

**Fig. S15. Phase behaviour of the ball-stick octagon system at high pressures. a** The caloric curves for the solid-solid phase transition at different pressures. For the two-step melting of the rotator crystal phase, we use the translational function **(b)** to identify the solid-hexatic phase transition and the caloric curve **(c)** to characterize the hexatic-liquid phase transition.

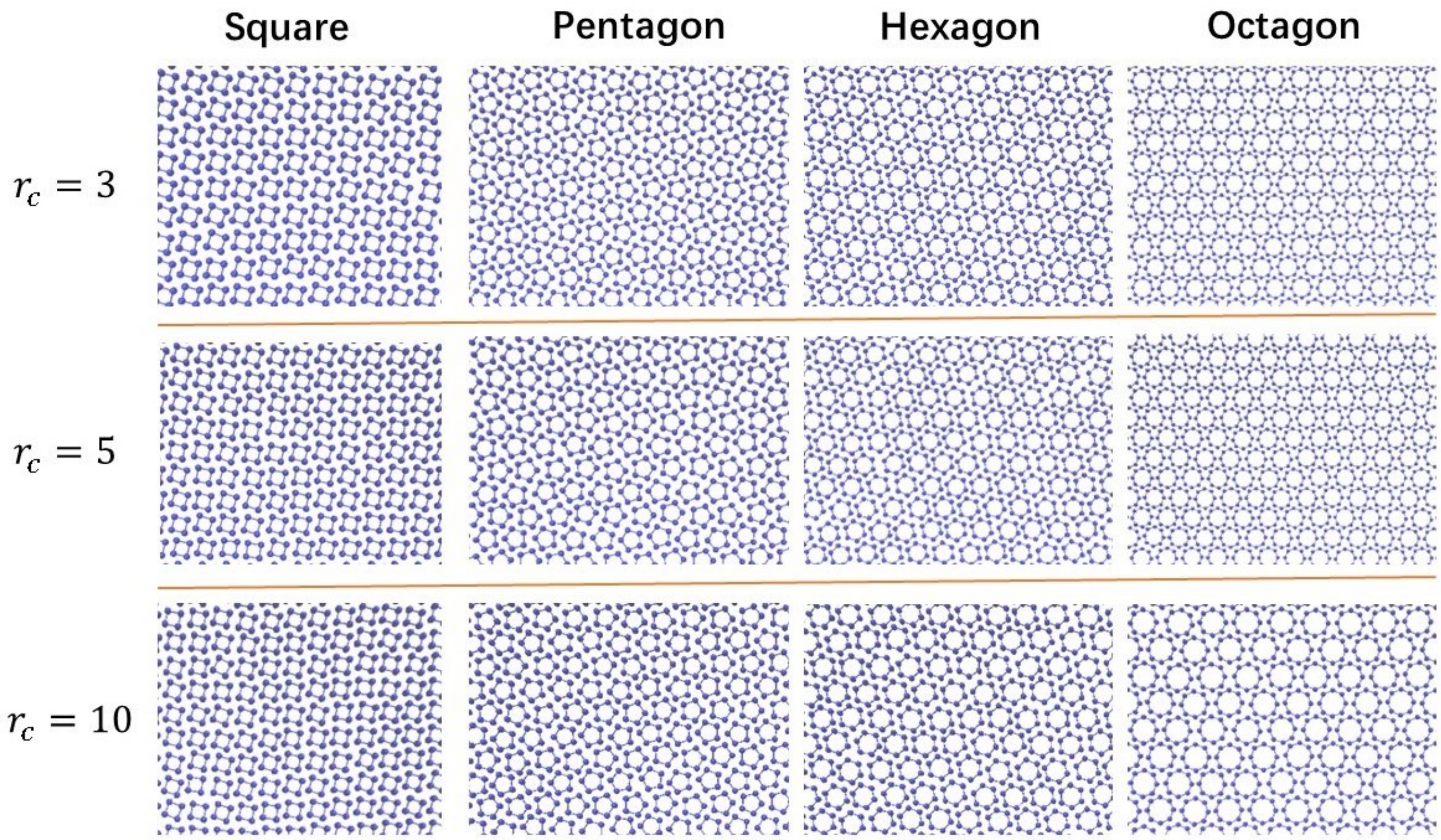

**Fig. S16 Packing modes of polygons with various $r_c$ values, exhibiting no qualitative differences.**